\documentclass[12pt, notitlepage]{article}
\pdfoutput=1

\usepackage{float}
\usepackage{subfig}
\usepackage{graphicx}
\usepackage[T1]{fontenc}
\usepackage[utf8]{inputenc}
\usepackage{authblk}
\usepackage{amsmath,bm}
\usepackage[left=3cm,right=3cm,top=2cm,bottom=2cm]{geometry}
\renewcommand\footnotemark{}
\begin{document}


\title{Exploring the new phase transition of CDT}         

\author{D.N.~Coumbe$^{a}$, J.~Gizbert-Studnicki$^{b}$, J. Jurkiewicz$^{b}$}

\affil{\small{\emph{$^a$The Niels Bohr Institute, Copenhagen University,\\
  Blegdamsvej 17, DK-2100 Copenhagen, Denmark.}}\footnote{E-mail: Daniel.Coumbe@nbi.ku.dk}}

\affil{\small{\emph{$^b$Faculty of Physics, Astronomy and Applied Computer Science, Jagiellonian University, ul.~prof. Stanislawa Lojasiewicza 11, Krakow, PL 30-348}}\footnote{E-mail: jakub.gizbert-studnicki@uj.edu.pl, jerzy.jurkiewicz@uj.edu.pl}}

\date{\small({Dated: \today})}          
\maketitle


\begin{abstract}
\noindent This work focuses on the newly discovered bifurcation phase transition of CDT quantum gravity. We define various order parameters and investigate which is most suitable to study this transition in numerical simulations.
  By analyzing the behaviour of the order parameters we present evidence that the transition separating the bifurcation phase and the physical phase of CDT is likely a second or higher-order transition, a result that may have important implications for the continuum limit of CDT.

\vspace{1cm}

\end{abstract}


\begin{section}{Introduction}

Assuming only key aspects of quantum mechanics and general relativity, and including few additional ingredients, causal dynamical triangulations (CDTs) define a particularly simple approach to quantum gravity. The simplicity of construction and plenitude of results has made CDT a serious contender for a nonperturbative theory of quantum gravity. There now exists strong evidence that CDT has a classical limit that closely resembles general relativity on large distance scales~\cite{Ambjorn:2008wc}, while on short distance scales it has produced some exciting hints about the nature of spacetime at the Planck scale, including evidence that the number of spacetime dimensions may dynamically reduce~\cite{Ambjorn:2005db}, a result that has also been reported in numerous other approaches to quantum gravity \cite{Lauscher:2005qz,Horava:2009if,Modesto:2008jz,Atick:1988si}. 

CDT gives an approximate description of continuous spacetime via the connectivity of an ensemble of locally flat $n$-dimensional simplices. In order to reproduce general relativity in the classical limit it seems the introduction of a causality condition is a necessary requirement \cite{Ambjorn05}, such that the lattice can be foliated into spacelike hypersurfaces of fixed topology. By only including geometries in the path integral measure that admit such a foliation, the unphysical features observed in dynamical triangulations without a causality condition (see \cite{Catterall:1994pg,Bialas:1996wu,deBakker:1996zx} and more recently \cite{Coumbe:2014nea,Ambjorn:2013eha}) appear to be suppressed, yielding a semi-classical geometry that closely resembles general relativity.  

CDT discretises the continuous path integral into a partition function \cite{Regge:1961px}

\begin{equation} \label{eq:CDTPartitionFunction}
Z_{E}={\sum_{T}}\frac{1}{C_{T}}e^{-S_{EH}},
\end{equation}

\noindent and transforms the Einstein-Hilbert action into the discretised Einstein-Regge action

\begin{equation} \label{eq:GeneralEinstein-ReggeAction}
S^{Regge}_{E}=-\left(\kappa_{0}+6\Delta\right)N_{0}+\kappa_{4}\left(N_{4,1}+N_{3,2}\right)+\Delta\left(2N_{4,1}+N_{3,2}\right).
\end{equation}

\noindent Equation~(\ref{eq:CDTPartitionFunction}) is defined by the sum over all possible triangulations $T$, where $C_{T}$ is a symmetry factor dividing out the number of equivalent ways of labelling vertices in $T$. 

CDT defines two types of 4-dimensional triangulations, the $\left(4,1\right)$ and $\left(3,2\right)$ simplices (see Ref.~\cite{Ambjorn05} for more details). The number of $\left(4,1\right)$ simplices in Eq. (\ref{eq:GeneralEinstein-ReggeAction}) is given by $N_{4,1}$, the number of $\left(3,2\right)$ simplices is denoted by $N_{3,2}$ and the number of vertices in a triangulation is given by $N_0$. Equation (\ref{eq:GeneralEinstein-ReggeAction}) is a function of three bare coupling constants: $\kappa_{0}$, $\Delta$ and $\kappa_{4}$. $\kappa_{0}$ is inversely proportional to Newton's constant, $\Delta$ defines an asymmetry parameter quantifying the ratio of the length of space-like and time-like links on the lattice and $\kappa_{4}$ is related to the cosmological constant, and is typically tuned in numerical simulations to a (pseudo-)critical value. Fixing $\kappa_{4}$ in this way allows one to take an infinite-volume limit, leaving a two-dimensional parameter space spanned by the bare couplings $\kappa_{0}$ and $\Delta$.

\begin{figure}[h]
  \centering
  \includegraphics[width=0.8\linewidth,natwidth=610,natheight=642]{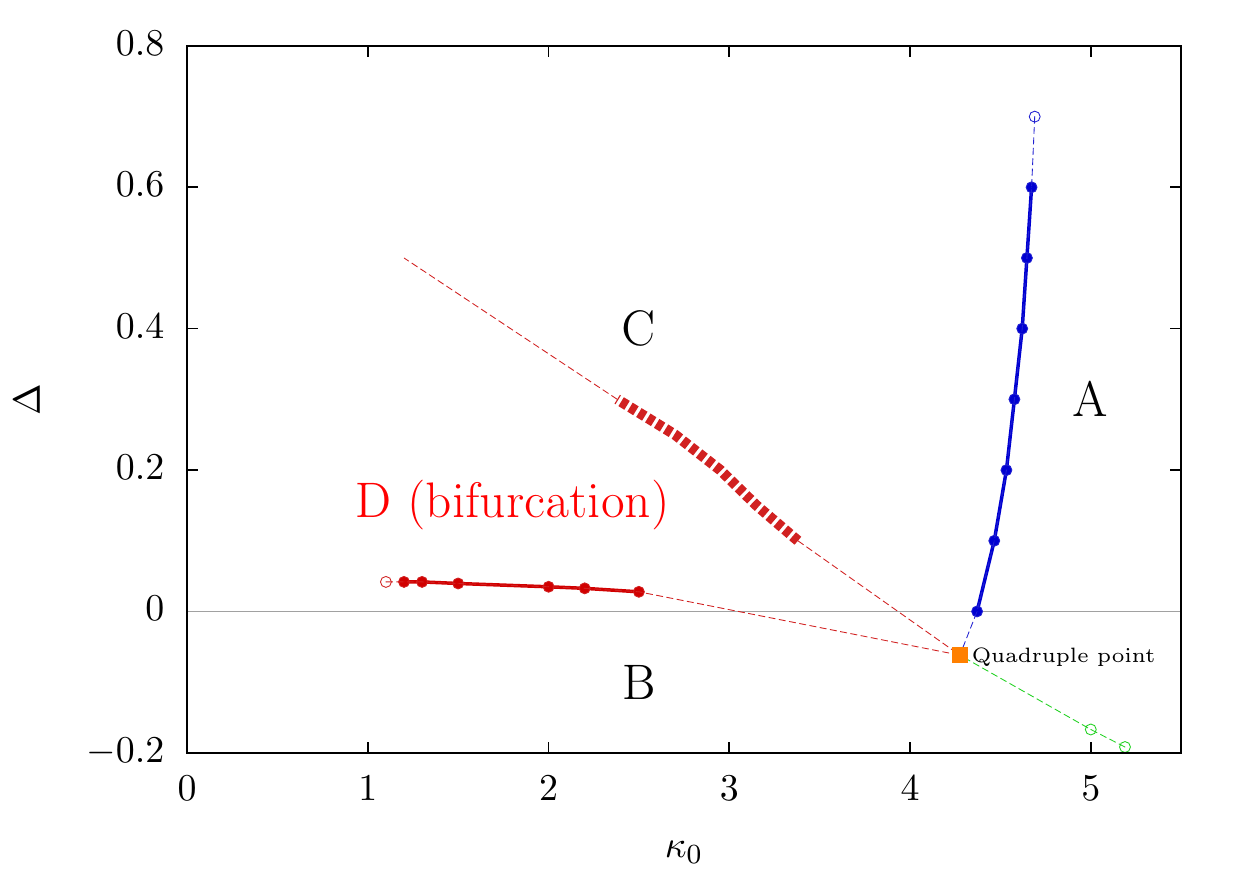}
  \caption{\small The phase diagram of 4-dimensional CDT. Filled points denote actual measurements while dashed lines represent extrapolations.}
\label{PD}
\end{figure}




The parameter space of CDT has now been mapped out in some detail, as shown in Fig. \ref{PD}, and consists of four distinct phases. Phases A and B are generally regarded as lattice artifacts containing unphysical geometric properties \cite{Ambjorn05}. Phase C, however, closely resembles 4-dimensional de Sitter space on large distance scales \cite{Ambjorn:2008wc}. The possibility of taking a continuum limit within phase C seemed a real possibility following the discovery of a second-order phase transition dividing phases B and C. However, the discovery of a fourth so-called bifurcation phase (D) existing between phases B and C makes it difficult or impossible to approach this second-order transition from within the physically interesting phase C. 
This motivates the need to investigate the location and order of the (C-D) bifurcation
 phase transition, since if the transition was second-order it would re-establish the possibility of taking a continuum limit in CDT.

\end{section}


\begin{section}{Defining an order parameter to study the bifurcation transition}\label{DefOP}

\begin{subsection}{Overview}

  In order to locate and study the critical behaviour of the transition dividing the bifurcation and de Sitter phases we seek an order parameter (OP) that is approximately zero in one phase and non-zero in the other. Hence, by taking the $n$th-order derivative of an appropriately defined order parameter one should in principle be able to determine the order of the transition. For example, in the infinite volume limit a first order transition is characterised by a discontinuity in the first order derivative at the transition point, whereas a continuous function should be observed for higher-order transitions. 
In numerical simulations one usually considers the susceptibility $\chi$ defined via the variance of the order parameter $\text{OP}$,
  
\begin{equation}\label{chi}
\chi_{\text{OP}}=\langle \text{OP}^{2} \rangle - \langle \text{OP} \rangle ^{2}.
\end{equation}

\noindent One then searches the parameter space for peaks in the susceptibility, whose presence would indicate the existence of a (pseudo-)critical point. By measuring how the position of such points changes with increasing volume one can in principle determine the location of the transition in the infinite volume limit via extrapolation. Critical exponents can also be determined using the same method, thereby helping to determine the order of the transition. 

It is important to carefully define a suitable order parameter. A good order parameter should capture the true nature of the transition and provide a strong signal/noise ratio. We now investigate various order parameters to find one that gives the strongest signal of the bifurcation transition, and therefore is the most suitable
to measure its precise location and order. Order parameters analysed in this article  can be  divided into two major groups.  The first group comprises order parameters which capture only global features of CDT triangulations. Such global order parameters have already been proposed in Refs.~\cite{Ambjorn:2011cg, Ambjorn:2012ij}, where they were used to locate and analyse the previously discovered A-C and B-D transitions.\footnote{As we now know that phase D 
exists, the former "B-C" transition now becomes the B-D transition.} Examples of such global OPs include: $N_0$, $N_1$, $N_2$ and $N_4$ which denote the total number of vertices, links, triangles and 4-simplices in a triangulation, respectively.  We have analysed all of the above  OPs, finding similar qualitative behaviour. In the following sections we will focus on a particular combination, namely

\begin{equation}
\text{OP}_0=\text{conj}(\Delta) =  2N_{4,1}+N_{3,2} -6 N_0.
\label{EqOP0}
\end{equation}

\noindent In order to analyse the bifurcation transition we performed a series of measurements of this OP for a range of bare coupling constants that begin in  
phase D and end in phase C. We study a particular path within the phase diagram for which we fix $\kappa_0=2.2$ and vary $\Delta$. Therefore  $\text{OP}_0$ given by Eq.~(\ref{EqOP0}), which is conjugate to $\Delta$ in the bare CDT action~(\ref{eq:GeneralEinstein-ReggeAction}), seems to be a particularly good choice. The same order parameter was also used in Refs.~\cite{Ambjorn:2011cg, Ambjorn:2012ij} to analyse the former B-D$^1$ transition in a similar way.

The second group of order parameters focuses on  microscopic geometric properties of the underlying CDT triangulations. It was shown in Ref.~\cite{Ambjorn:2015qja} that  the distribution of volume in the bifurcation phase is markedly different than in phase C, with spatial volume concentrated in clusters connected by vertices of very large coordination number (the number of 4-simplices sharing a given vertex). This change of the geometric structure  can be exploited to signal the phase transition. Inside the bifurcation phase both the average scalar curvature $\bar R(t)= 2\pi \frac{N_0(t)}{N_3(t)} - \text{C}$ (where $C=6 \, \arccos(1/3) - 2\pi > 0$) and the maximal coordination number of a vertex $O\big(v(t)\big)$ differ significantly between spatial slices of odd and even time $t$, whereas there is no such difference in phase C. One can quantify this difference by defining the order parameters \cite{Ambjorn:2015qja}

\begin{equation}\label{EqOP1}
OP_1 = \left| \bar R(t_0) - \bar R( t_0+1) \right|
\end{equation}

\noindent and

\begin{equation}\label{EqOP2}
OP_2 = \Big|  \text{max}\big[O\big(v(t_0\big) \big] - \text{max}\big[O\big(v(t_0+1\big) \big] \Big|,
\end{equation}

\noindent where the (integer) time $t_0$ is chosen to be the closest to the centre of volume of a triangulation.\footnote{In our approach the discrete centre of volume $t_0$ is defined up to one time slice, therefore to calculate $\text{OP}_1$ and $\text{OP}_2$ we first choose $t_0$ and measure 3 values of $OP$ for $t_0-1$, $t_0$ and $t_0+1$ and then choose the highest one.} A detailed analysis of all three order parameters is presented in section~\ref{results}.

\end{subsection}

\begin{subsection}{Thermalization and error estimates}

  When performing Monte Carlo simulations it is important to ensure the lattice is thermalized before beginning to take measurements. Ensuring thermalization is particularly important to this work as we aim to explore phase transition lines that are typically associated with very long auto-correlation lengths. For each of our measurement series we performed thermalization checks by dividing the data series into two sets and statistically comparing them. An example of such a check is presented in Fig. \ref{FigThermal} where we plot  the $\text{OP}_2$ order parameter defined in Eq. (\ref{EqOP2}) as a function of Monte Carlo time (proportional to the number of attempted moves). We check whether a given configuration range is thermalized by splitting the data set in two and comparing the average and standard deviation of each set. A comparison between the two data sets gives good statistical agreement, as shown in Fig.~\ref{FigThermal}. We find the longest autocorrelation lengths closest to the phase transition, and that the autocorrelation time increases with total volume. At the transition point, the order parameter tunnels between two metastable values, with the frequency of transition decreasing for larger total volumes (see~Fig. \ref{FigThermal} (right)). The statistical agreement between subsets of data for the larger volume ensemble is slightly worse than for the smaller ensemble because for the same physical simulation period we observe fewer metastable transitions, meaning local variations have had less time to average out.\footnote{For the 160k ensemble we only observe two metastable transitions over the entire simulation period of almost nine months, so the average transition period is around three months.}





\begin{figure}[h]
  \centering
  \scalebox{.6}{\includegraphics{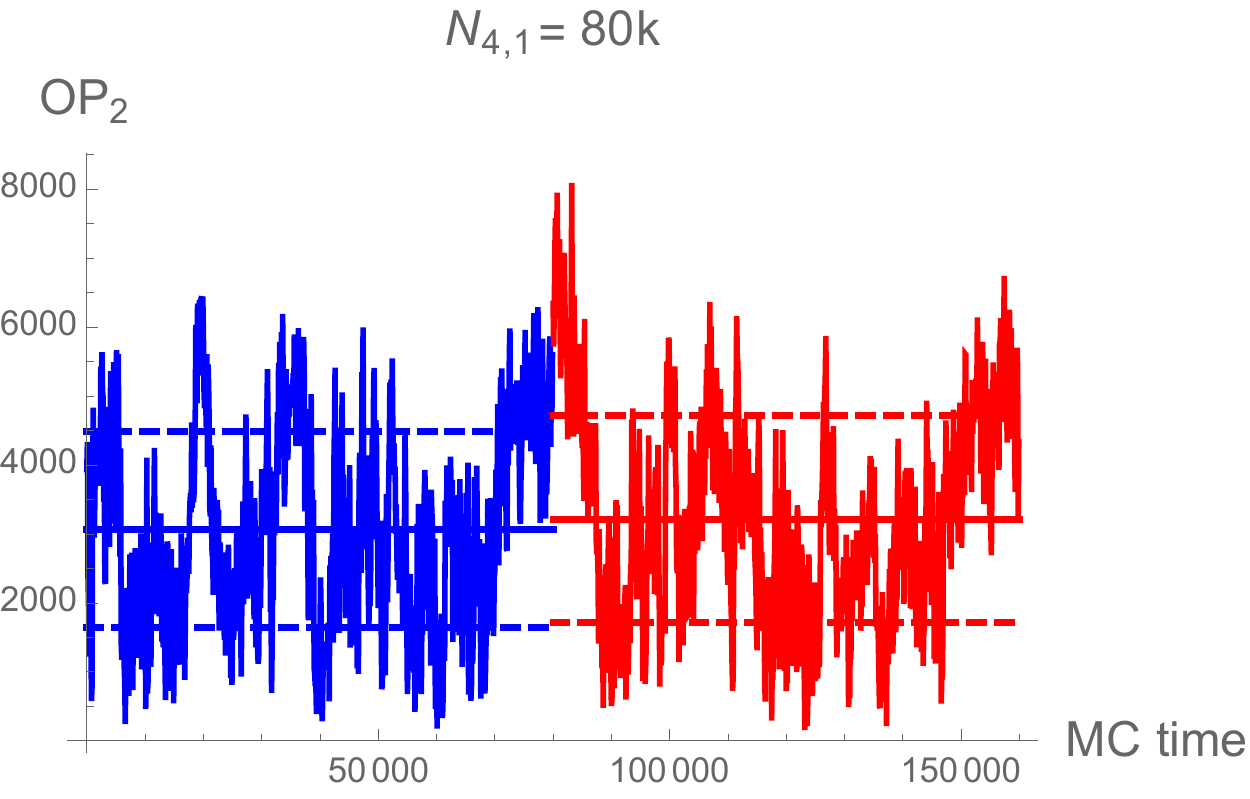}}
  \scalebox{.6}{\includegraphics{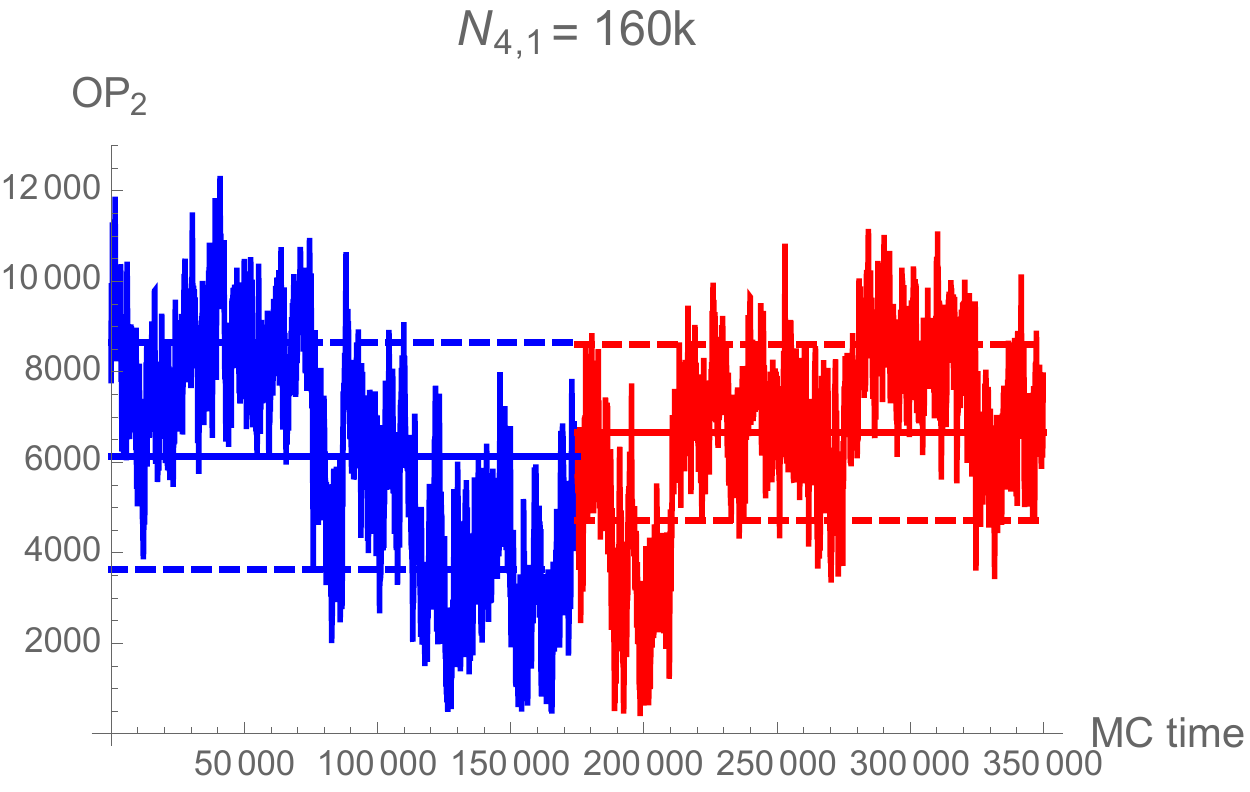}}
  \caption{\small An example thermalization check based on the $\text{OP}_2$ order parameter. The order parameter is plotted as a function of simulation time (proportional to the number of attempted Monte Carlo moves) for our point closest to the phase transition and for lattice volumes of $N_{4,1}=80,000$ (left) and $N_{4,1}=160,000$ (right), respectively. The data is divided into two subsets (blue and red), whose statistical properties are compared. The mean value is denoted by a solid line and the dashed lines indicate $\pm$1 standard deviation error bounds.}
\label{FigThermal}
\end{figure}

When performing Monte Carlo simulations it is also important to accurately estimate sources of statistical errors. Statistical errors in this work are calculated using a single-elimination (binned) jackknife procedure, after blocking the data to account for autocorrelation errors. When autocorrelation errors are important the statistical error increases with increasing block size, and when autocorrelation errors are insignificant the error is largely independent of block size. For this reason we calculate the associated error for various block sizes, selecting the block size for which the statistical error is maximised.  An example of such a procedure is presented in Fig.~\ref{FigError} where we plot the error in the measurement of the susceptibility $\chi_{\text{OP}_2}$ at the point closest to the phase transition. The error is estimated by a jackknife procedure for each block size and is plotted as a function of the number of blocks. The error typically increases with the number of blocks, eventually stabilising around a constant, as shown for the smaller volume ensemble presented in Fig. \ref{FigError} (left). In some cases the largest error is observed for a small number of blocks, which appears to be the case for the larger volume ensemble close to the phase transition point (see Fig. \ref{FigError} (right)). As already discussed, for this empirical data we observe only two 
metastable transitions in the order parameter over the entire simulation period, this likely means the jackknife procedure is overestimating the error. We adopt a cautious attitude and take the highest value as our error estimate.  




\begin{figure}[h]
  \centering
  \scalebox{.6}{\includegraphics{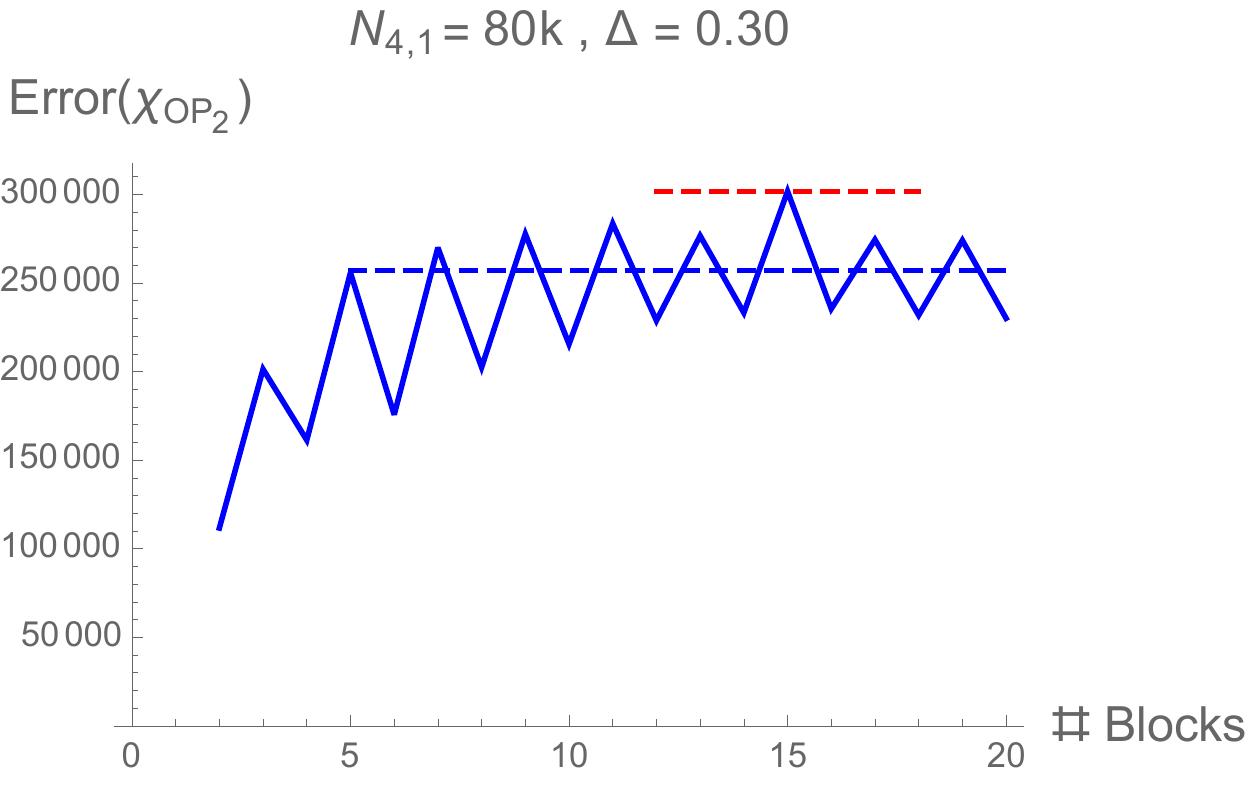}}
  \scalebox{.6}{\includegraphics{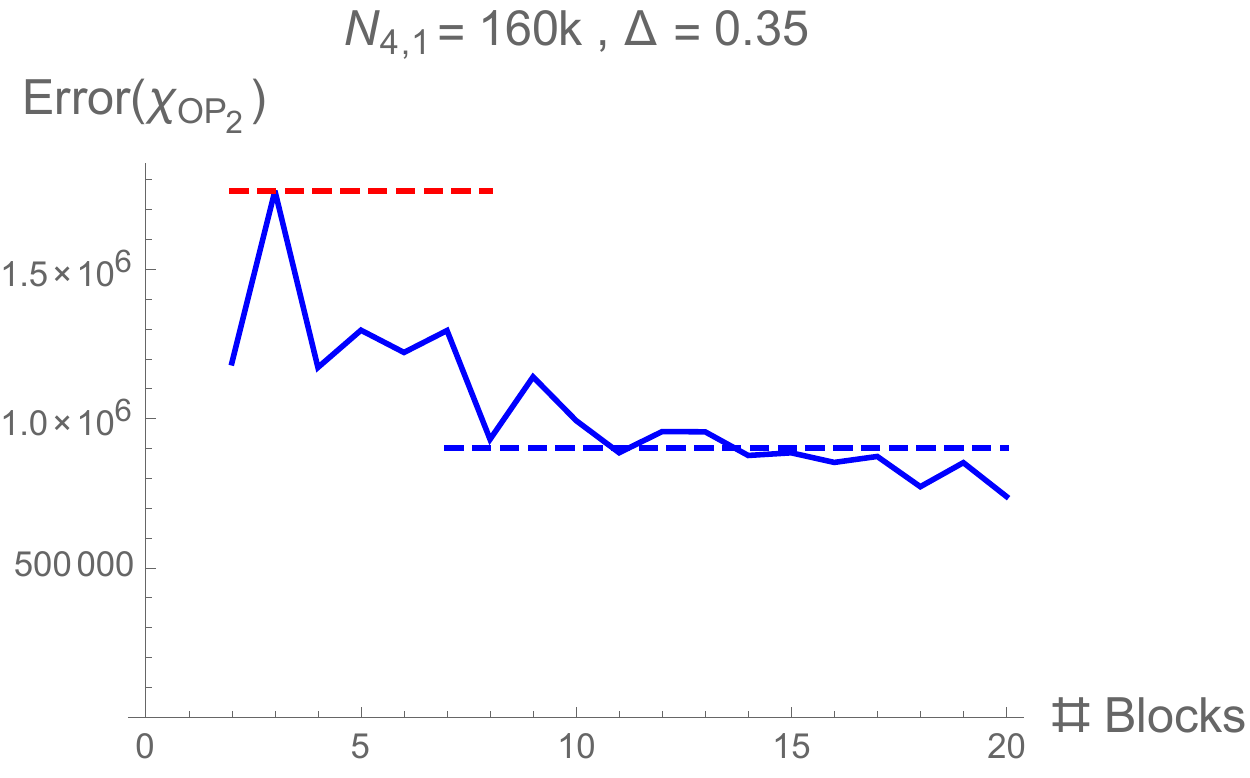}}
  \caption{\small The statistical error of the susceptibility $\chi_{\text{OP}_2}$ calculated for the point closest to the phase transition and for lattice volumes of $N_{4,1}=80,000$ (left) and $N_{4,1}=160,000$ (right), respectively. The data set is divided into blocks of identical size and then a single-elimination (binned) jackknife procedure is used to determine the statistical error. The size of the error depends on the number of blocks. We take the largest value (red dashed line) as our final error estimate.}
\label{FigError}
\end{figure}




\end{subsection}

\begin{figure}[H]
  \centering
  \scalebox{.4}{\includegraphics{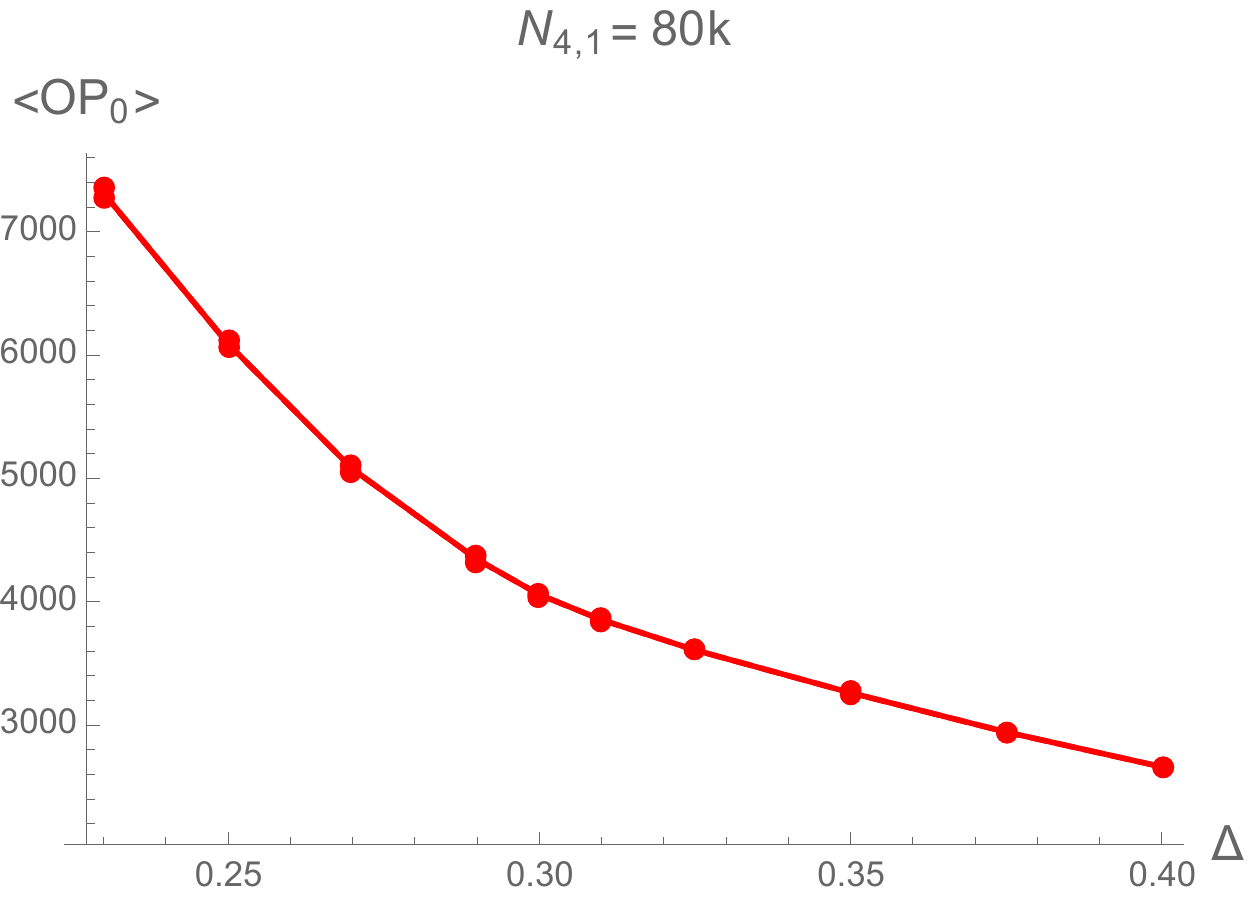}}
  \scalebox{.4}{\includegraphics{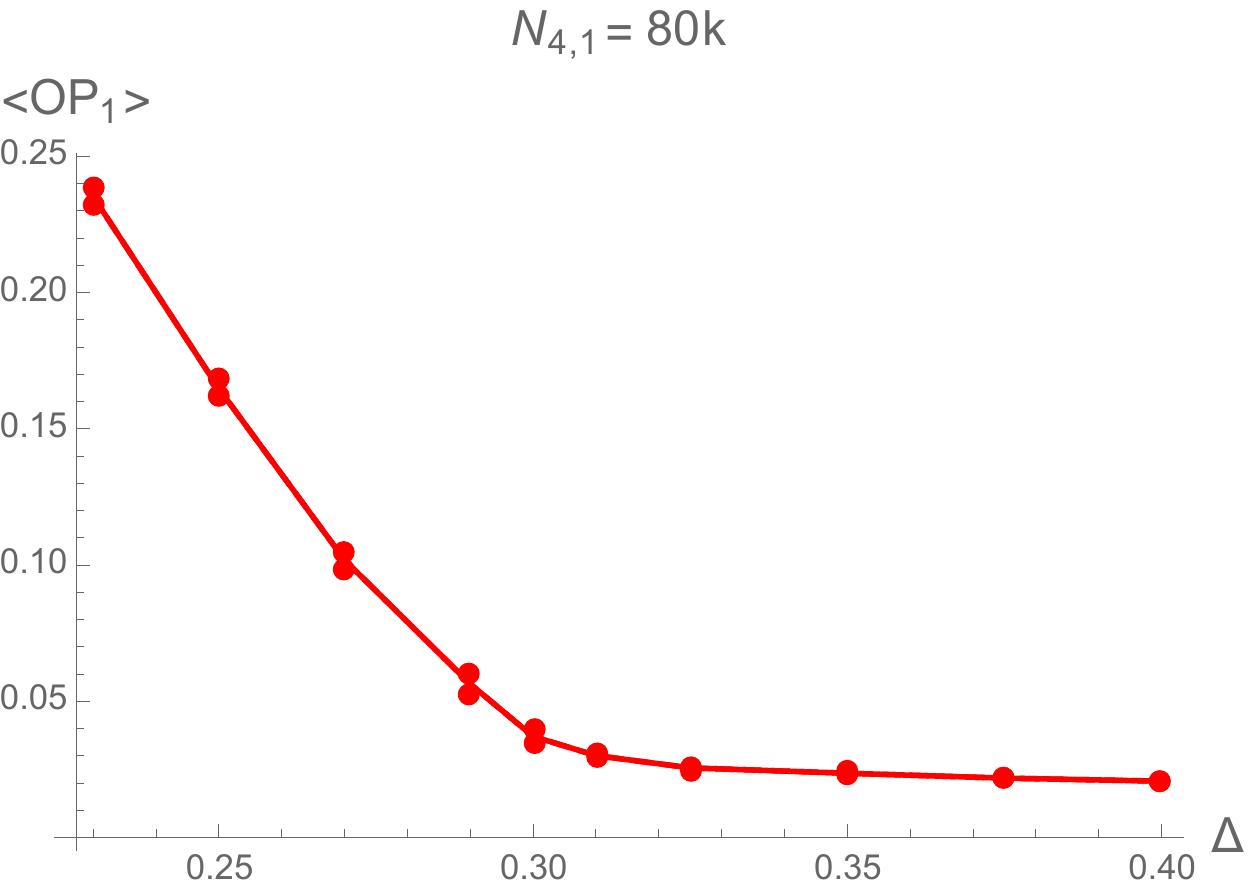}}
  \scalebox{.4}{\includegraphics{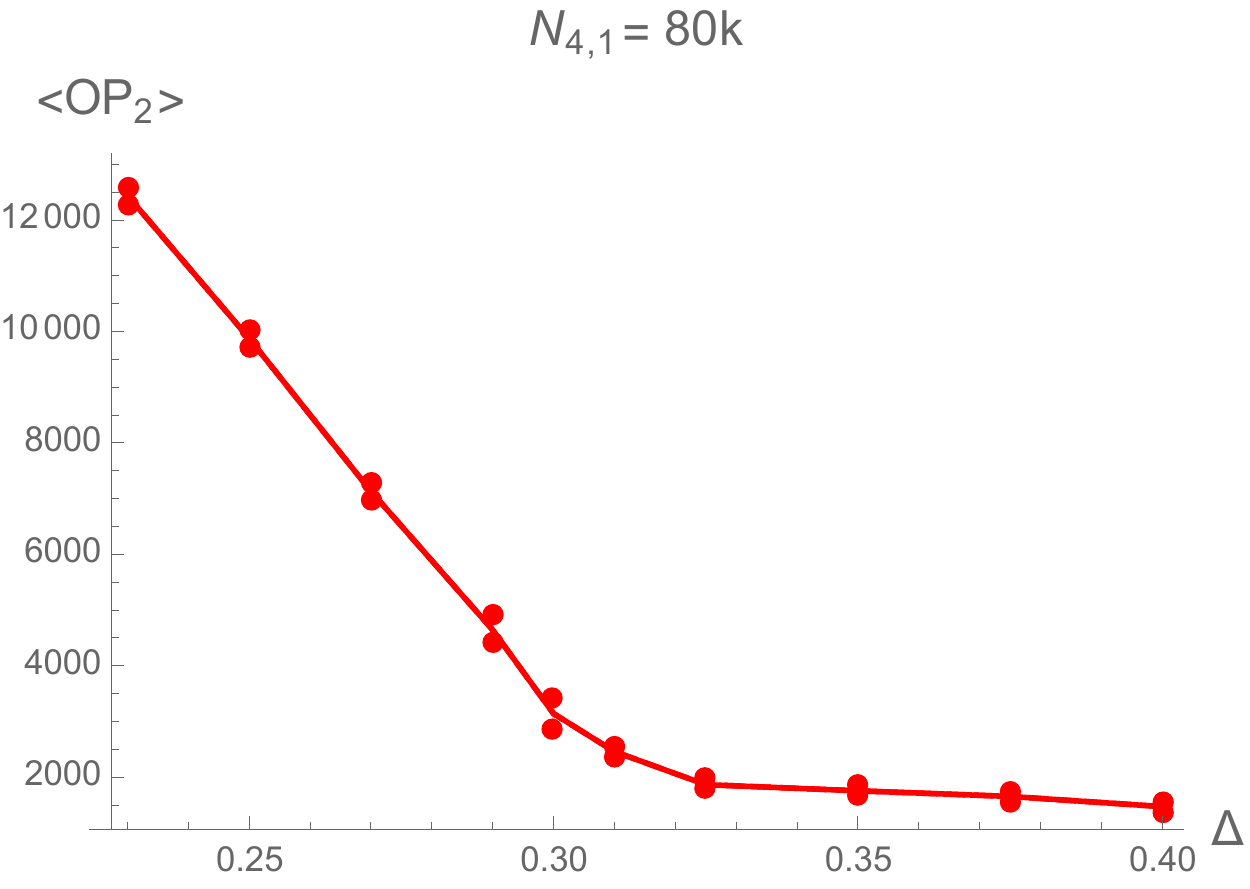}}
  \scalebox{.4}{\includegraphics{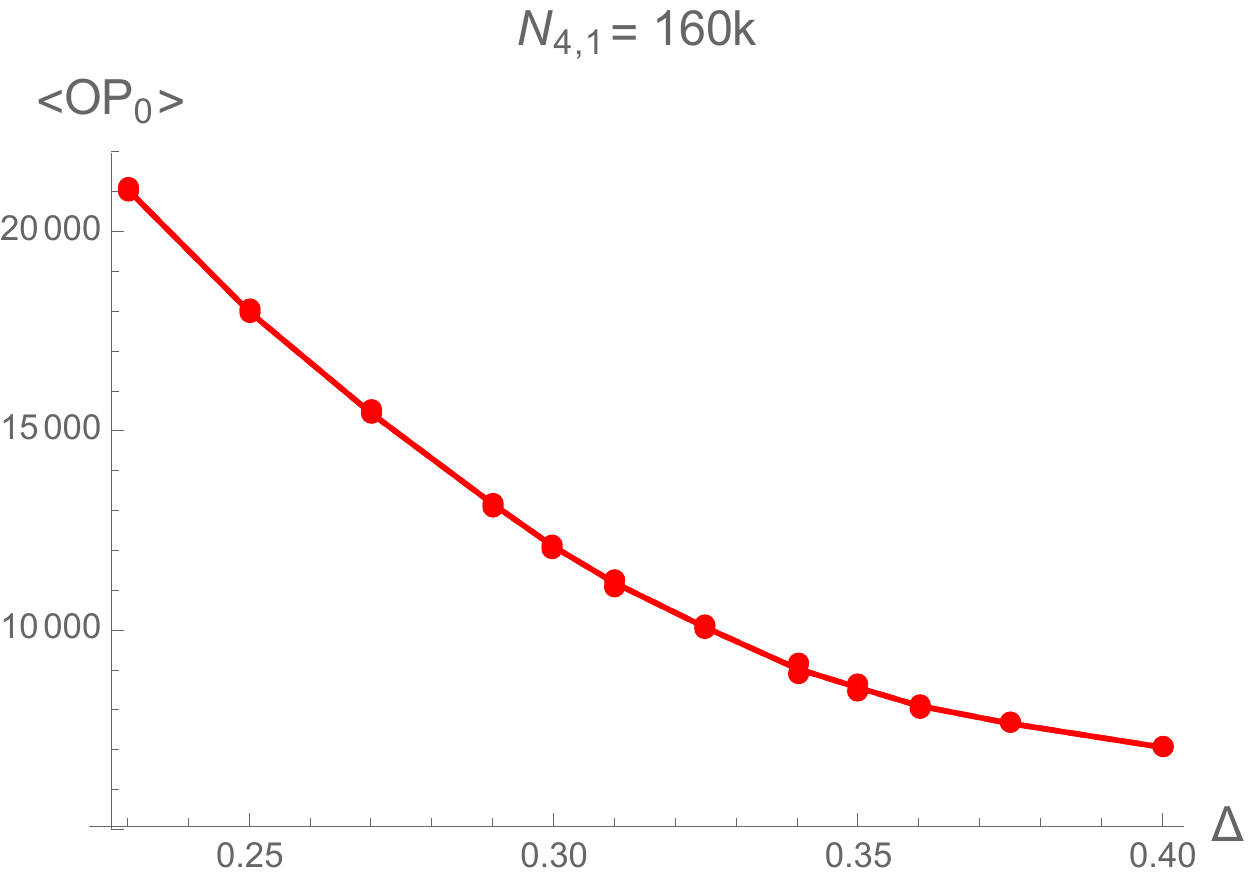}}
  \scalebox{.4}{\includegraphics{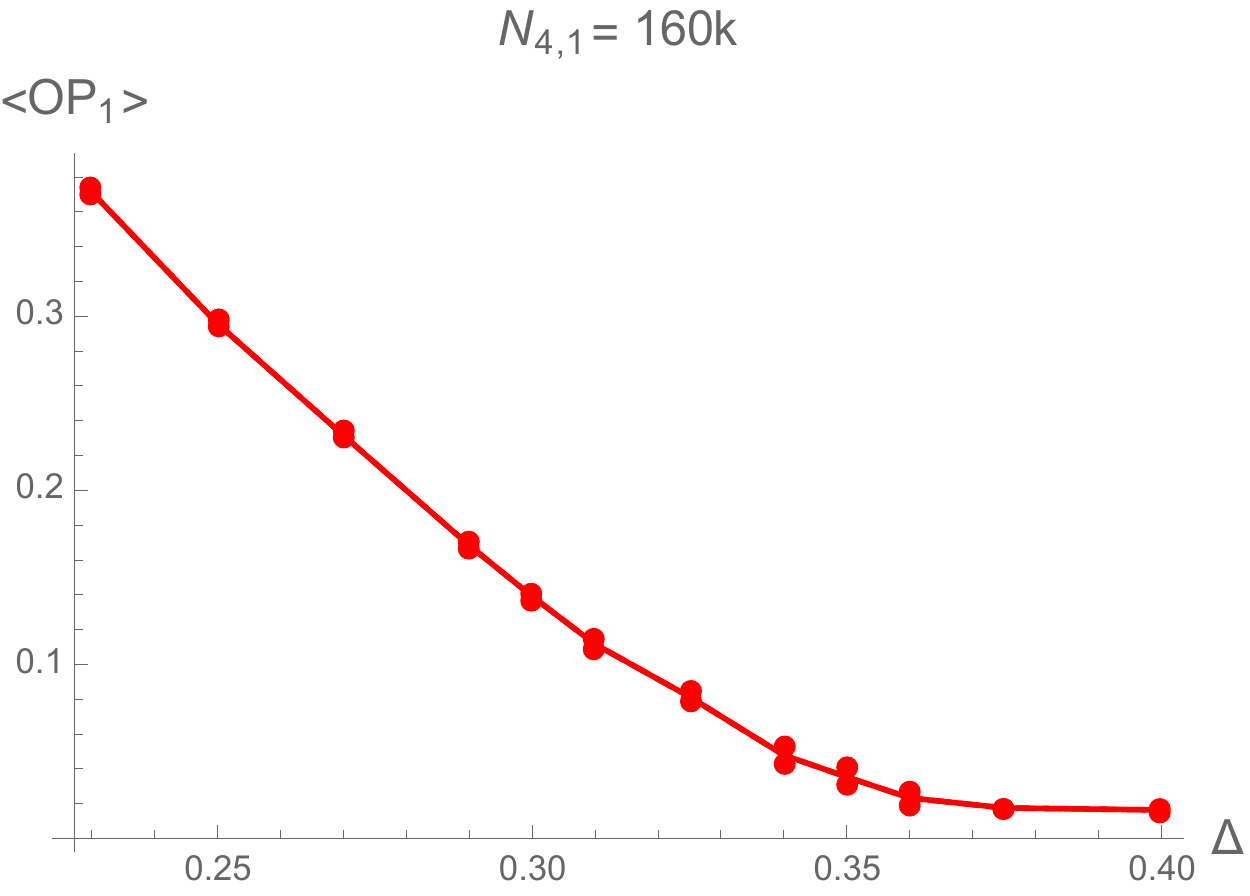}}
  \scalebox{.4}{\includegraphics{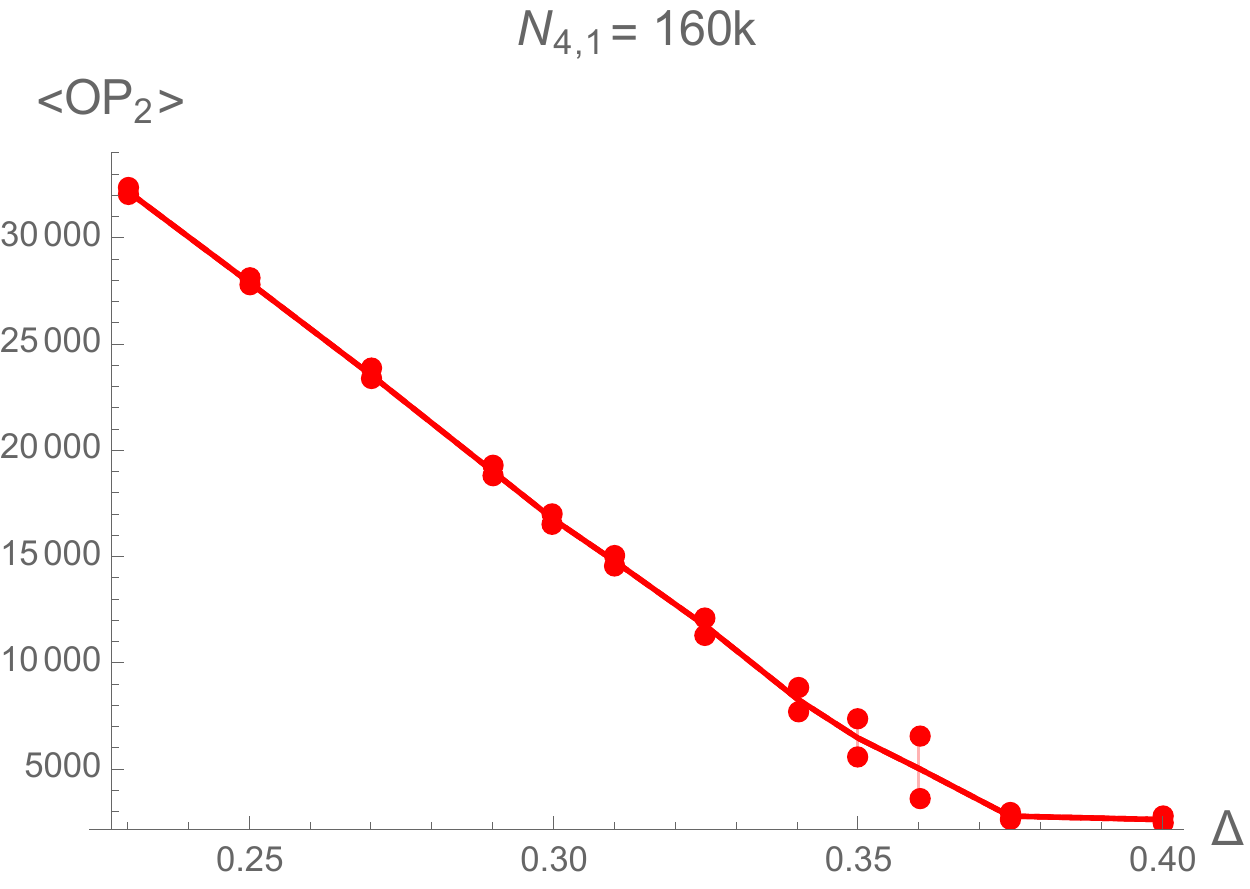}}
  \caption{\small The mean value $\langle\text{OP}\rangle$ as a function of $\Delta$ for three different order parameters $\text{OP}_0$ (left), $\text{OP}_1$ (centre) and $\text{OP}_2$ (right) and for two different lattice volumes $N_{4,1}=80,000$ (top) and $N_{4,1}=160,000$ (bottom). $\text{OP}_1$ and $\text{OP}_2$ both clearly change around $\Delta=0.27-0.325$ and $\Delta=0.325-0.375$ for $N_{4,1}=80,000$ and $N_{4,1}=160,000$, respectively, suggesting a phase transition. However, there is no clear signal of a transition when using $\text{OP}_0$.}
\label{FigOPav}
\end{figure}

\begin{subsection}{Results}\label{results}

We now present the results of our order parameter studies. We focus on three order parameters defined in section~\ref{DefOP}. Fig.~\ref{FigOPav} shows the mean value of the order parameters $\langle\text{OP}\rangle$ plotted as a function of $\Delta$ for fixed $\kappa_{0}=2.2$ and for two different lattice volumes $N_{4,1}=80,000$ and $N_{4,1}=160,000$. One clearly sees that all order parameters tend to zero (or a constant) for large $\Delta$ (inside phase C) and increase in value for smaller $\Delta$ (inside 
phase D). A clear change in behaviour of $\text{OP}_1$ and $\text{OP}_2$ can be seen around $\Delta=0.27-0.325$ and $\Delta=0.325-0.375$ for systems with $80,000$ and $160,000$ simplices of type (4,1), respectively, whereas there is no clear signal of the transition using the parameter $\text{OP}_0$.

In Fig.~\ref{FigOPX} we plot the susceptibility $\chi_{\text{OP}}$ of each order parameter defined in Eq. (\ref{chi}). A clear signal of the phase transition is observed only for the $\text{OP}_2$, where one can see a peak of susceptibility at the (pseudo-)critical points $\Delta^{crit}(80\text{k})=0.30\pm0.01$ and $\Delta^{crit}(160\text{k})=0.35\pm 0.01$. Interestingly, if one plots the ratio $\chi_{\text{OP}} / \langle\text{OP}\rangle$ one can also observe the transition peaks  using  $\text{OP}_1$ (see Fig. \ref{FigOPXav}). 
\begin{figure}[H]
  \centering
  \scalebox{.25}{\includegraphics{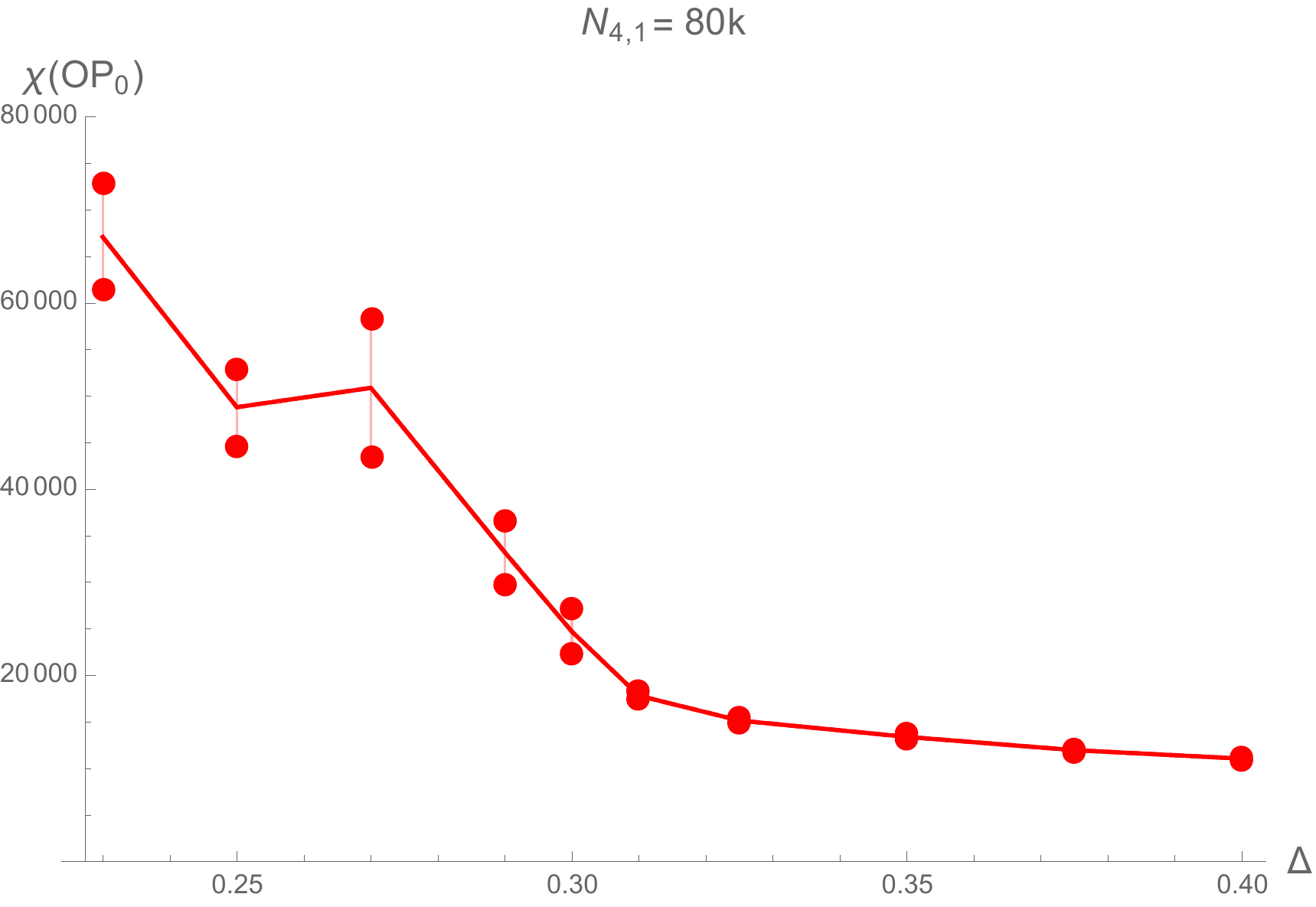}}
  \scalebox{.35}{\includegraphics{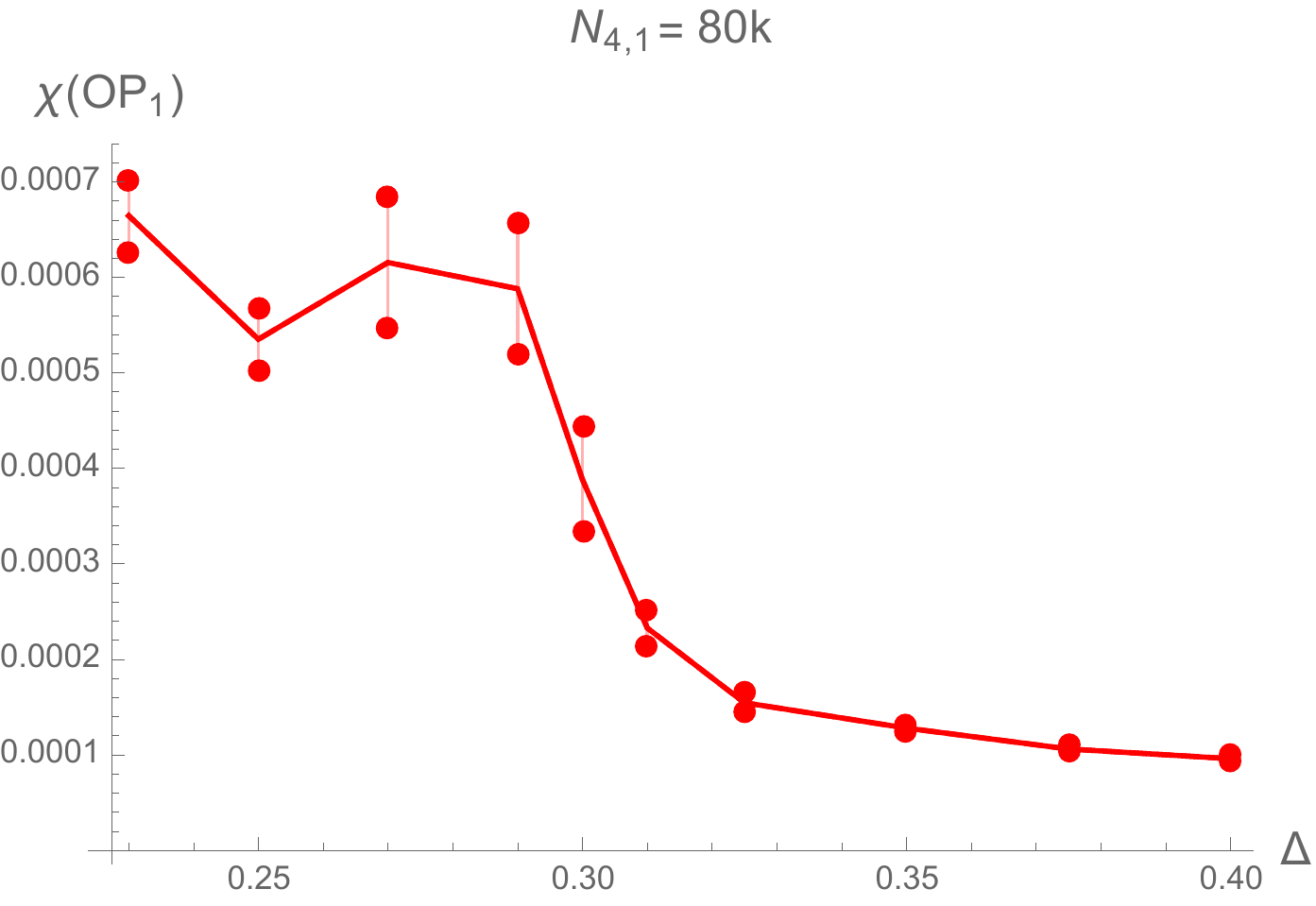}}
  \scalebox{.35}{\includegraphics{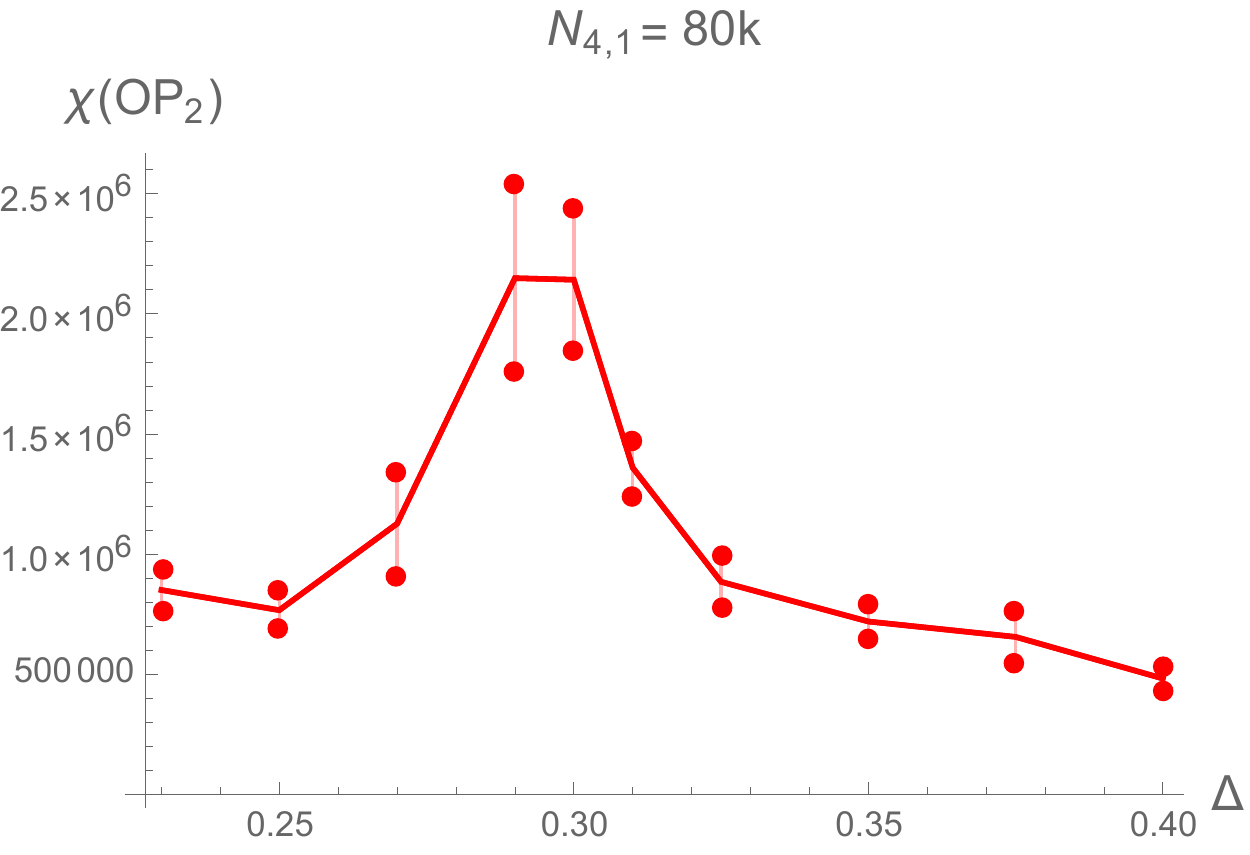}}
  \scalebox{.25}{\includegraphics{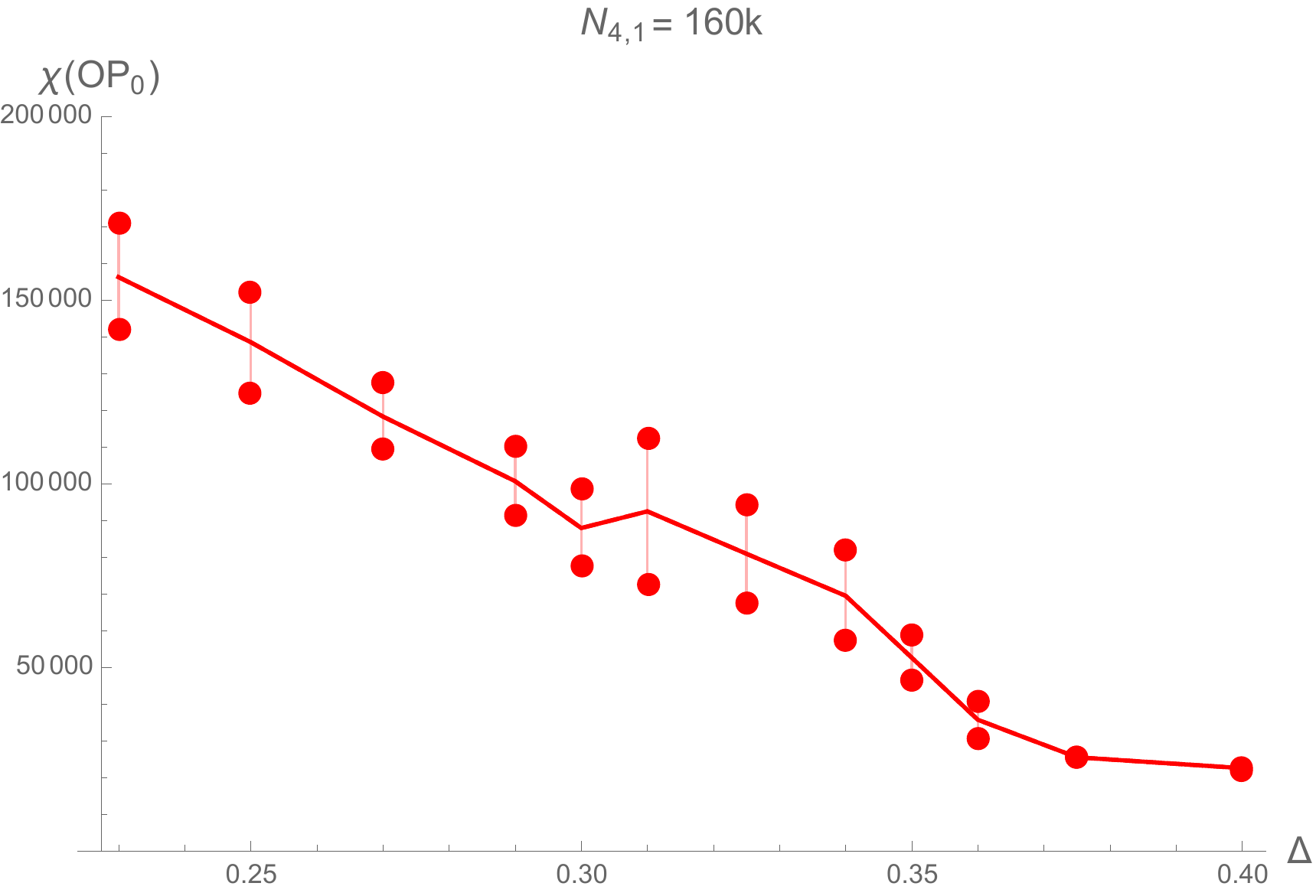}}
  \scalebox{.35}{\includegraphics{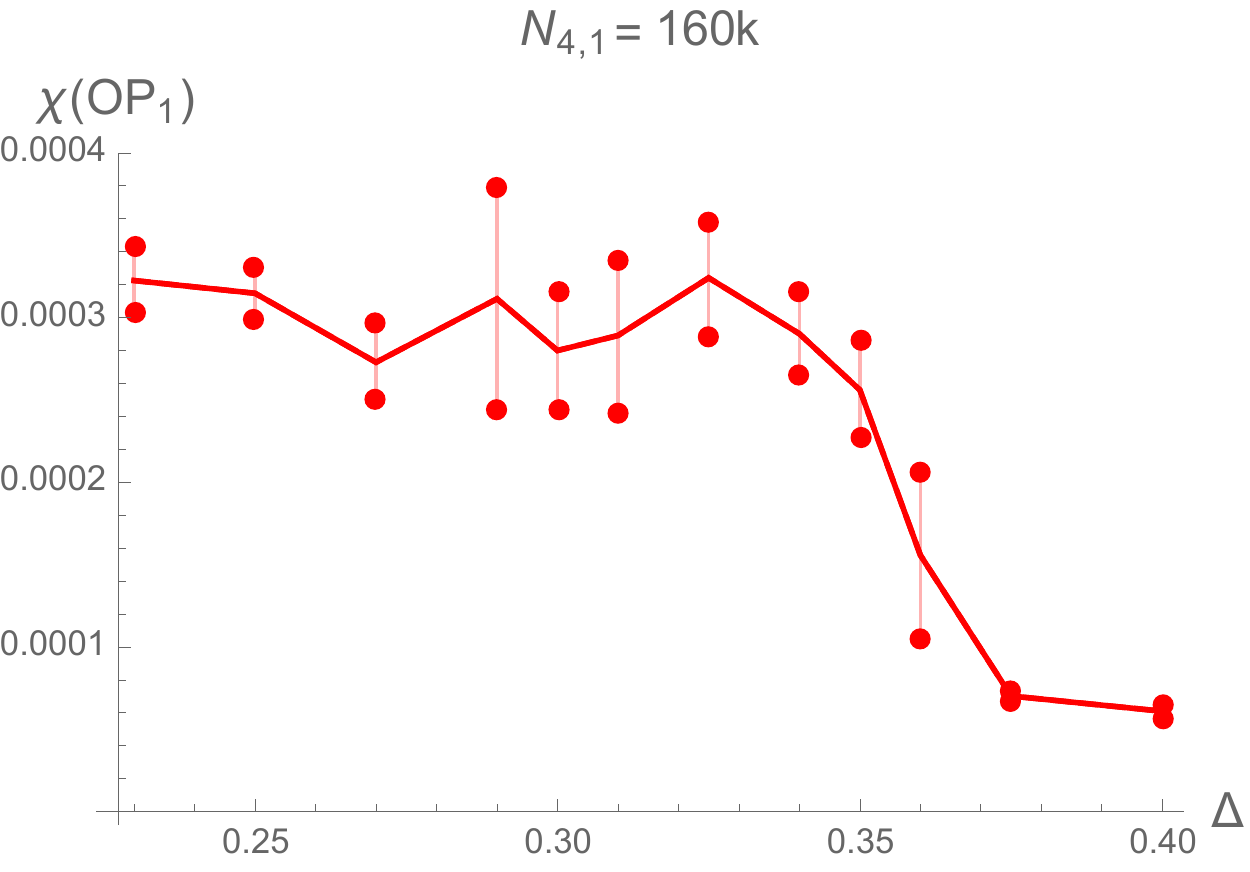}}
  \scalebox{.35}{\includegraphics{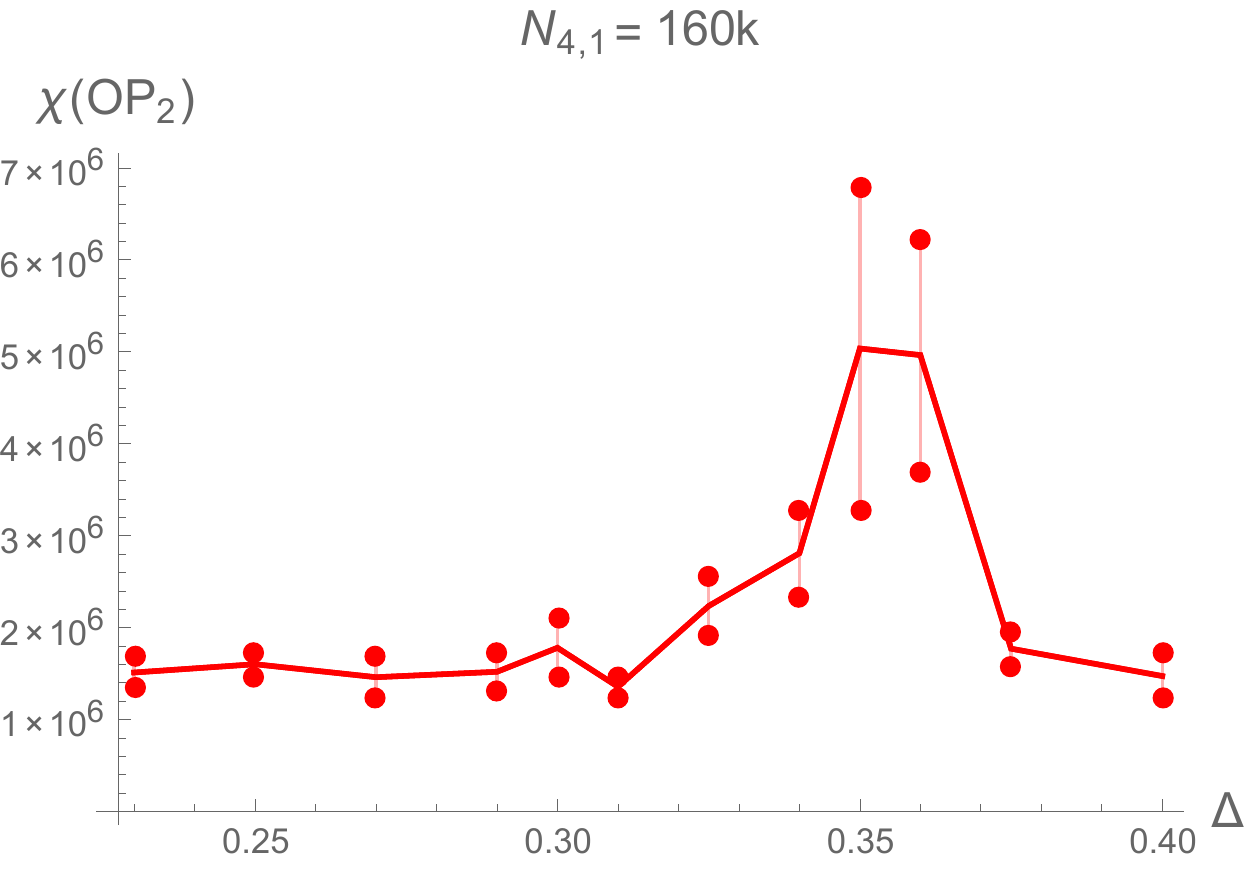}}
  \caption{\small The susceptibility $\chi_{\text{OP}}$  as a function of $\Delta$ measured for three different order parameters: $\text{OP}_0$ (left), $\text{OP}_1$ (centre) and $\text{OP}_2$ (right), and for two  different lattice volumes $N_{4,1}=80,000$ (top) and $N_{4,1}=160,000$ (bottom). The (pseudo-)critical $\Delta$ value at which the bifurcation transition occurs appears to be at $\Delta^{crit}=0.30\pm 0.01$ for $N_{4,1}=80,000$ and at $\Delta^{crit}=0.35\pm 0.01$ for $N_{4,1}=160,000$, as determined using $\text{OP}_2$.}
\label{FigOPX}
\end{figure}

\begin{figure}[H]
  \centering
  \scalebox{.35}{\includegraphics{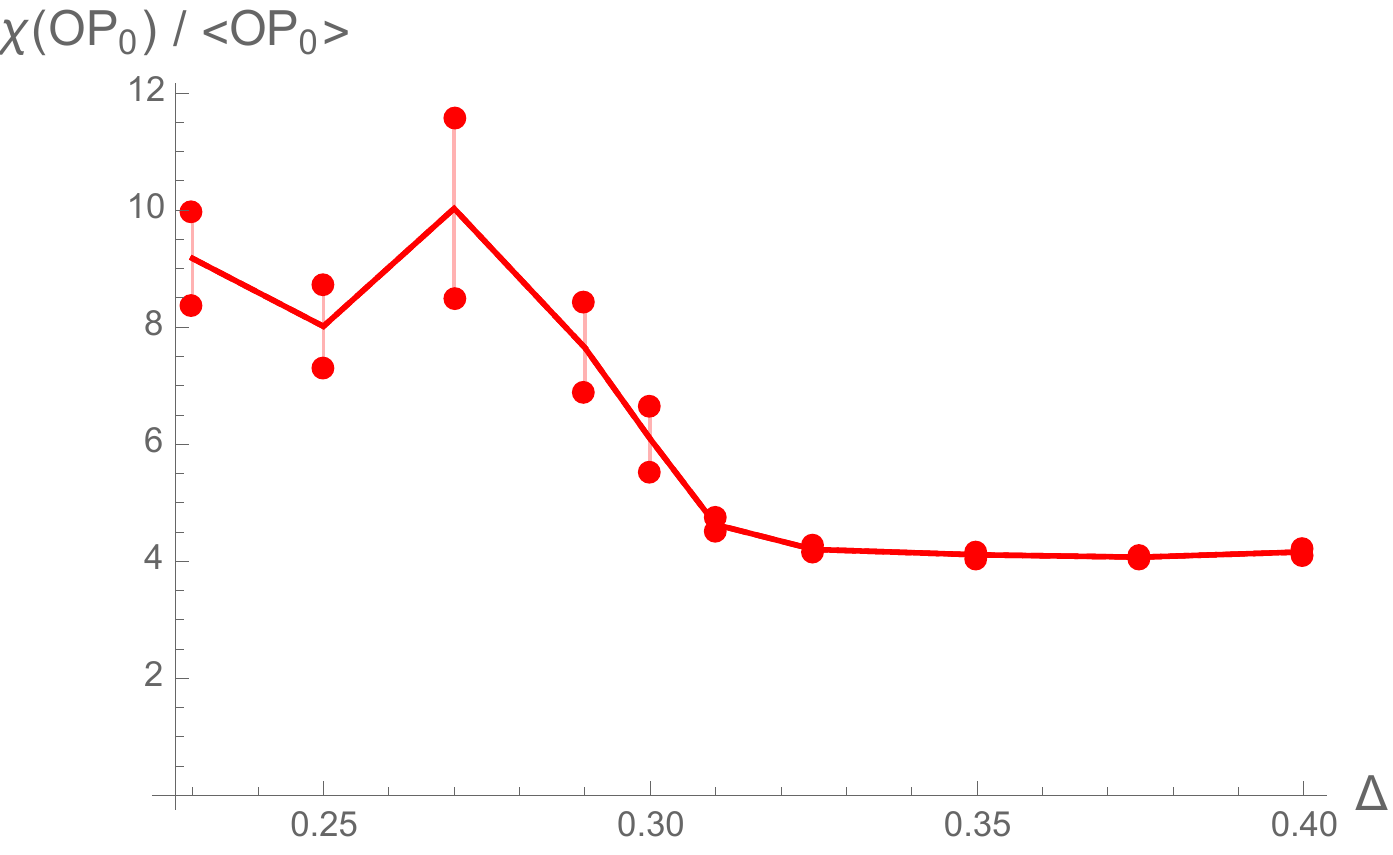}}
  \scalebox{.35}{\includegraphics{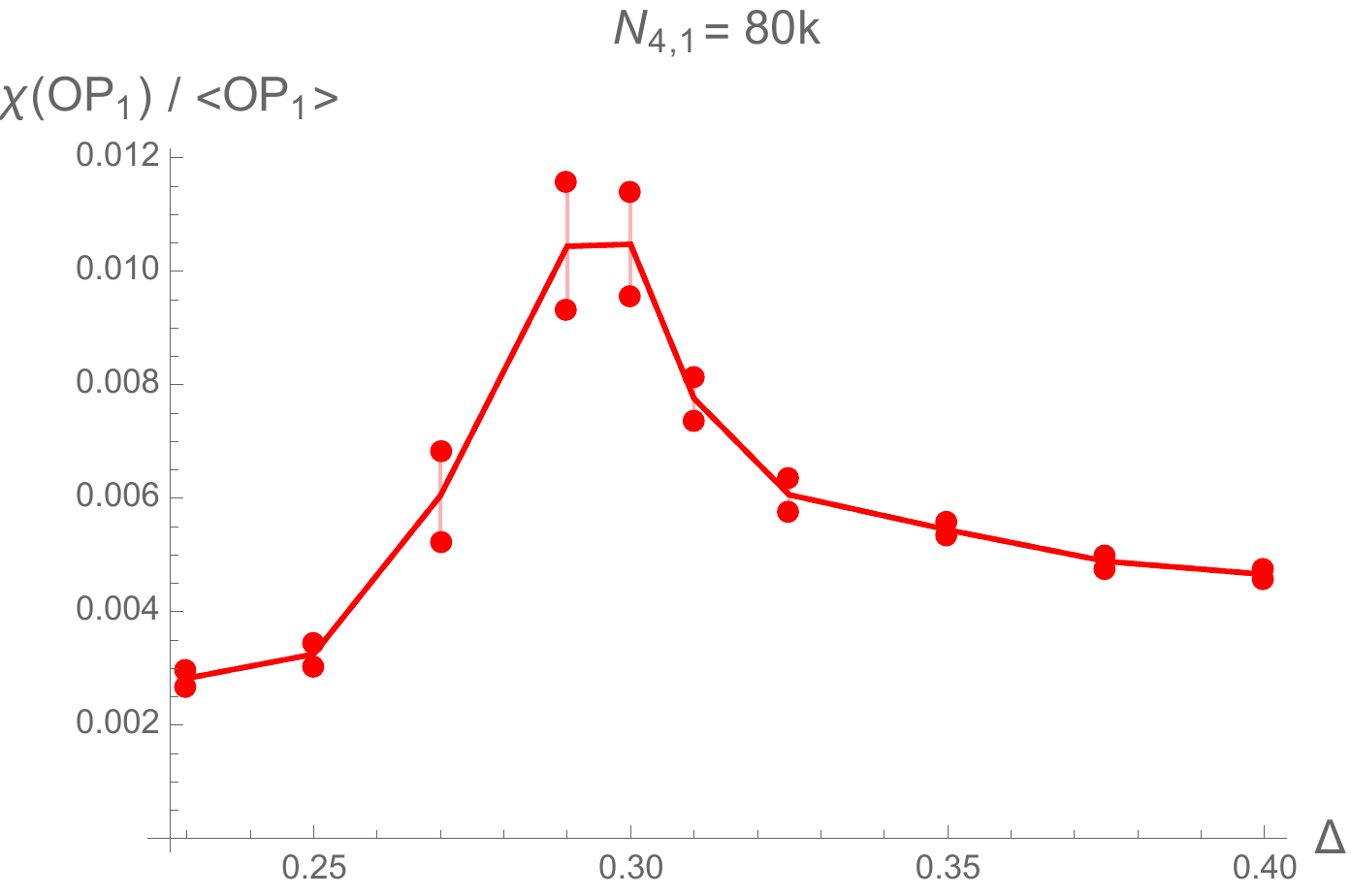}}
  \scalebox{.35}{\includegraphics{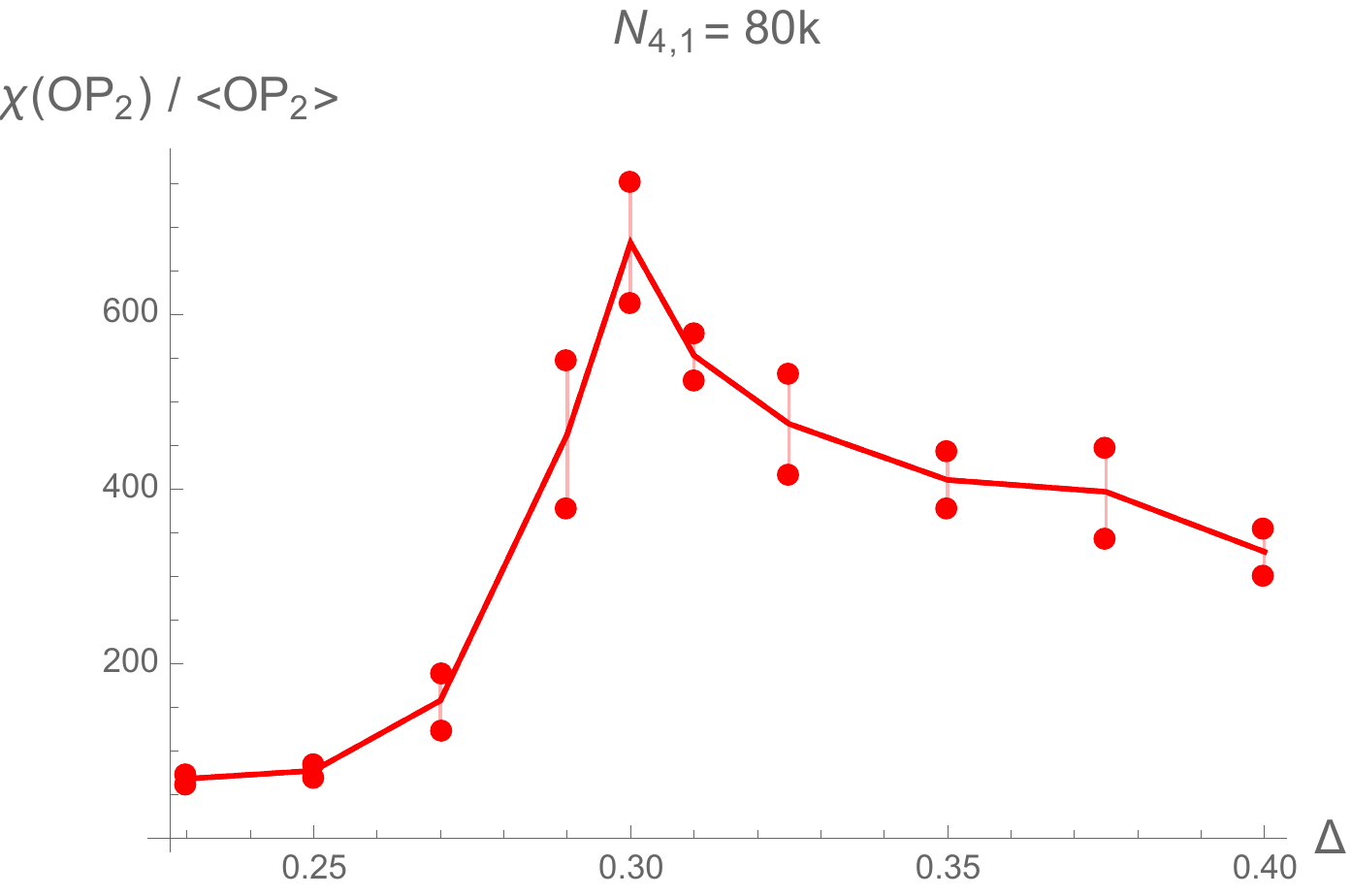}}
  \scalebox{.35}{\includegraphics{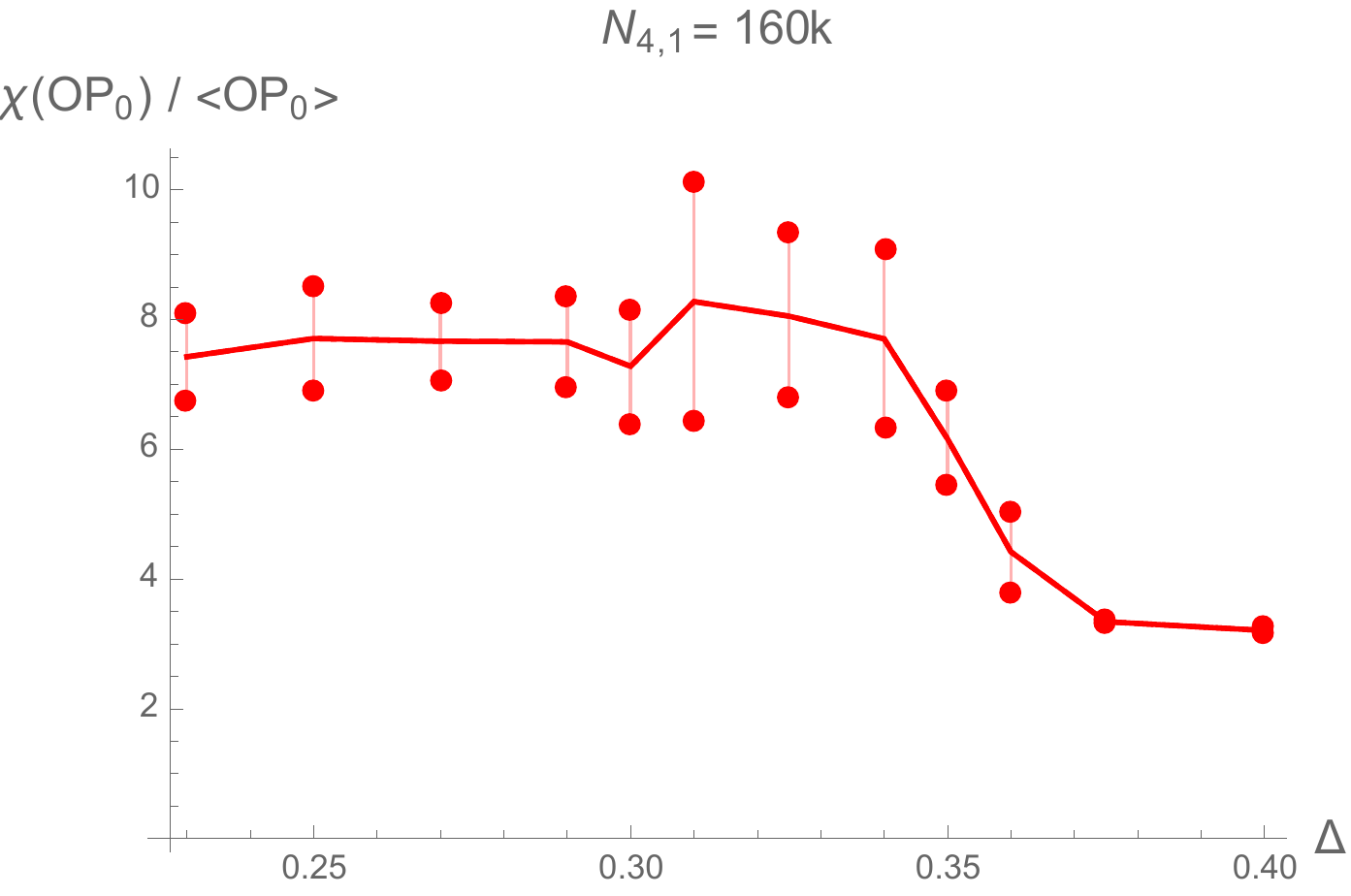}}
  \scalebox{.35}{\includegraphics{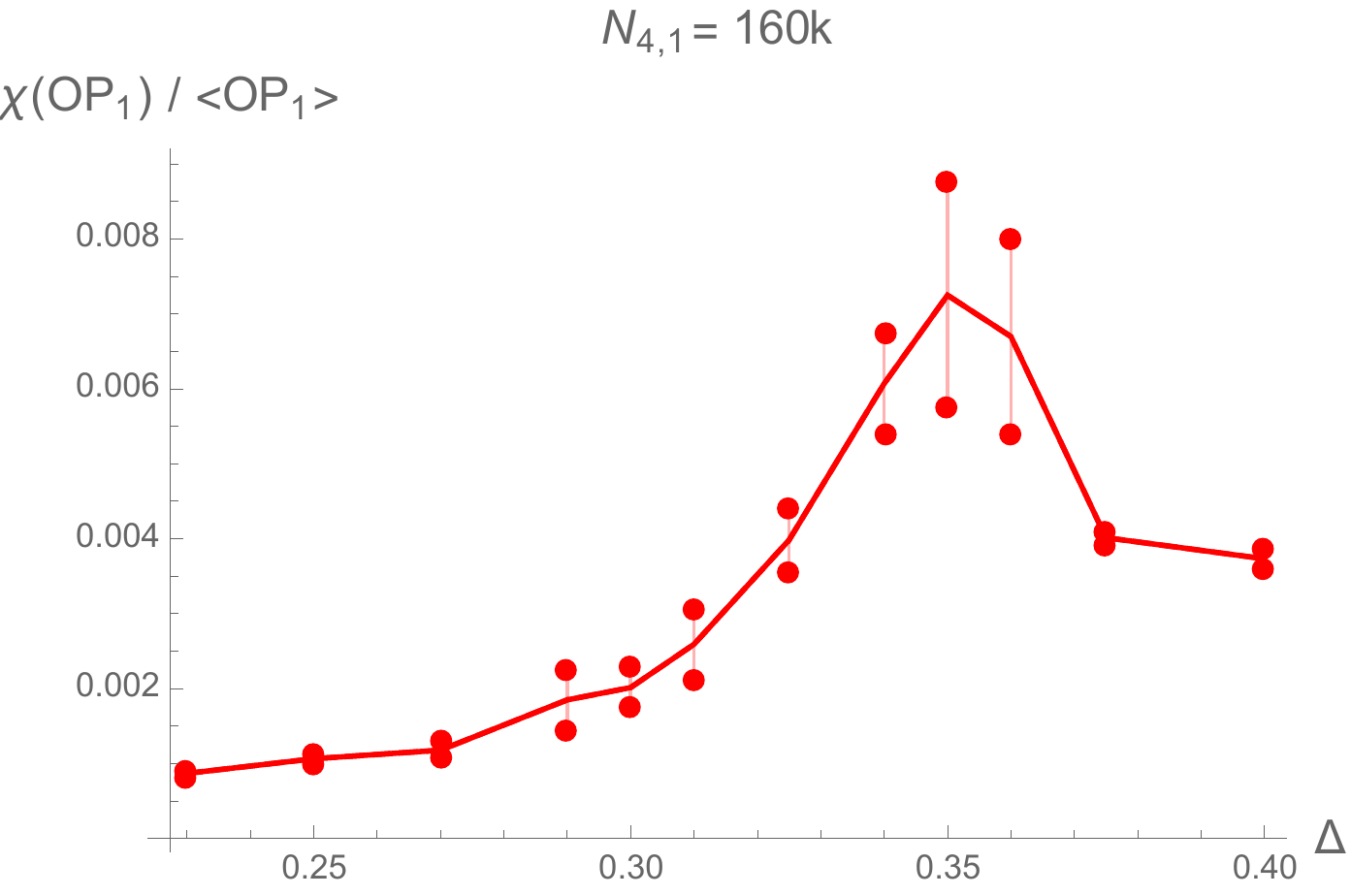}}
  \scalebox{.35}{\includegraphics{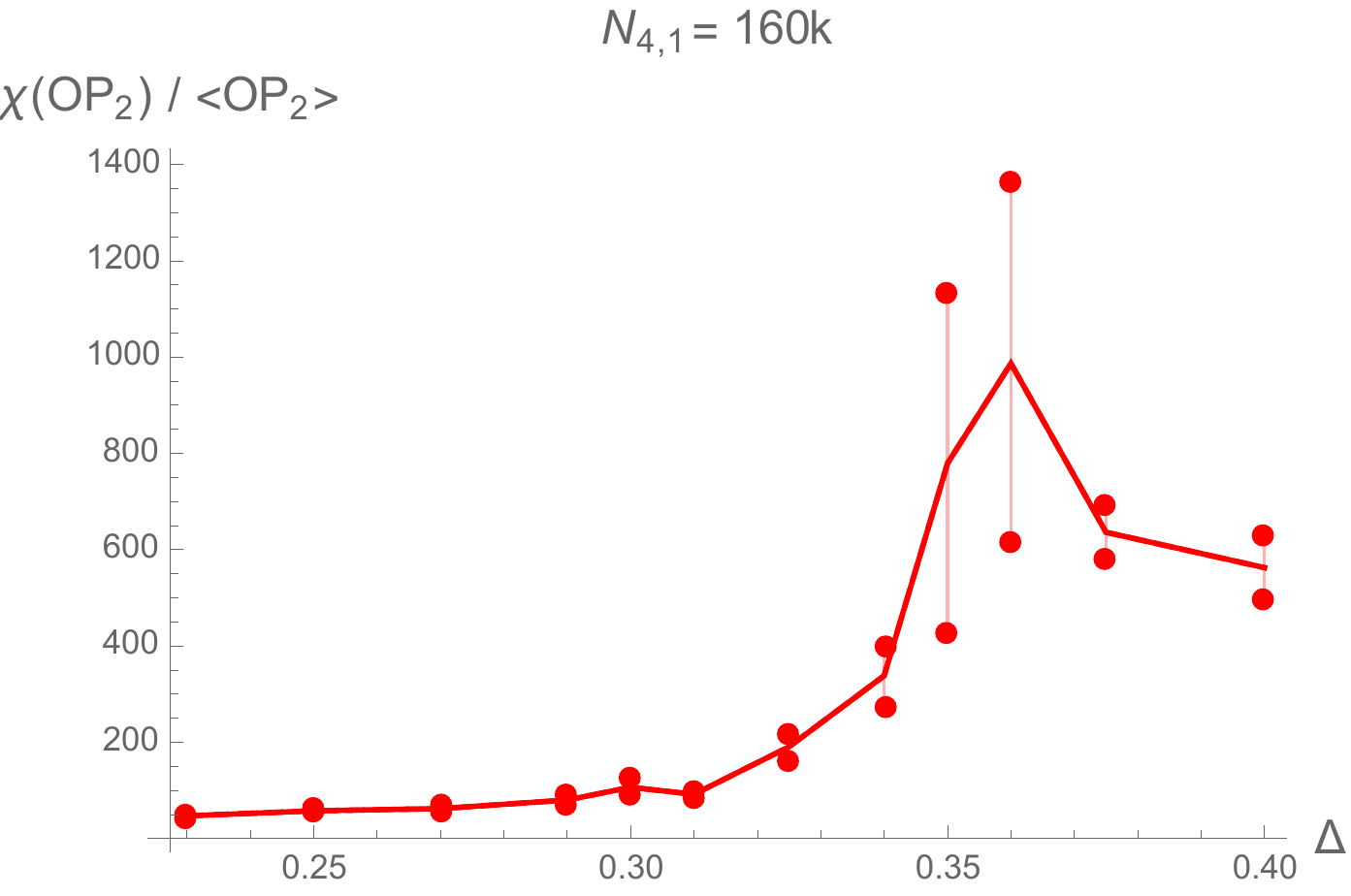}}
  \caption{\small The ratio $\chi_{\text{OP}} / \langle \text{OP} \rangle$ as a function of $\Delta$ measured for three different order parameters $\text{OP}_0$ (left), $\text{OP}_1$ (centre) and $\text{OP}_2$ (right) for two  different lattice volumes $N_{4,1}=80,000$ (top) and $N_{4,1}=160,000$ (bottom). The (pseudo-)critical $\Delta$ value at which the bifurcation transition occurs appears to be at $\Delta^{crit}=0.30\pm 0.01$ for $N_{4,1}=80,000$ and at $\Delta^{crit}=0.35\pm 0.01$ for $N_{4,1}=160,000$, as determined via the order parameters $\text{OP}_1$ and $\text{OP}_2$.}
\label{FigOPXav}
\end{figure}

The above results indicate that for the bifurcation transition the details of the geometry play an important role, and therefore order parameters based solely on global properties of the triangulation do not capture these details. The central difference between phase C and  phase D is related to the formation of periodic clusters of volume around singular vertices, which form a kind of tube structure (see Ref. \cite{Ambjorn:2015qja} for details). Such a structure does not exist in phase C, but is a generic property of 
phase D. Therefore, in order to observe the phase transition it is important to analyse the microscopic simplicial geometry. Even order parameters such as $\text{OP}_1$ only capture general features of the geometry (i.e. the difference in average curvature for different time slices), and are therefore not capable of capturing the microscopic details of the phase transition. This simple observation explains why the existence of the bifurcation phase went unnoticed during previous phase transition studies.


\end{subsection}
 \end{section}

\begin{section}{Discussion and Outlook}



Starting from a point in the parameter space with good semi-classical features, the hope is that one can establish a continuum limit by approaching a second order transition, thereby defining a smooth interpolation between the low and high energy regimes of CDT. The infinite correlation length associated with such a transition should allow one to shrink the lattice spacing to zero while keeping observables fixed in physical units. Such a continuous transition has been shown to exist in the CDT parameter space~\cite{Ambjorn:2011cg, Ambjorn:2012ij} and was originally thought to divide the semi-classical phase C from phase B. However, recent results~\cite{Ambjorn:2014mra, Ambjorn:2015qja} show that a new bifurcation phase (D) exists between phases C and B, which may prevent the possibility of taking a continuum limit from within phase C. Analysing the new transition between phases C and D is therefore very important, since a second order transition would re-establish the possibility of defining a continuum limit. To study this transition one must define an order parameter which signals the transition. In this article we have analysed two groups of order parameters, related to general and detailed features of the CDT simplicial geometries, respectively. We have shown that the parameters from the first (general) group, which were used in previous phase transition studies, do not work well with the new phase transition. However, the second (detailed) group of order parameters give a clear transition signal. Among the numerous order parameters tested, the strongest transition signal was given by $\text{OP}_2$, as defined by Eq. (\ref{EqOP2}).

The order of the new bifurcation transition remains an open question, although at least we now have an order parameter capable of determining it. It seems that the order parameter measured at the (pseudo-)critical point jumps between two different values (see Fig. \ref{FigThermal}) and that the frequency of such jumps  decreases with increasing volume. This result may suggest that the transition is first order. This is illustrated in Fig. \ref{FigHistogram} where we plot a histogram of the  $\text{OP}_2$ (normalised by the lattice volume) measured for two different volumes $N_{4,1}=80,000$ (blue) and $N_{4,1}=160,000$ (red), respectively. By fitting a double Gaussian function to the measured data we observe two clearly separated peaks.\footnote{As we are able to establish the phase transition point only with finite precision the height of the two peaks is different. The peaks would be the same height at the (pseudo-)critical point.} The peak separation is slightly smaller for a larger total volume. A similar situation was previously observed at the 'old' B-C (now called the B-D) phase transition (which is very likely second order)~\cite{Ambjorn:2012ij}, where the peak separation reduced with increasing volume. 

\begin{figure}[h]
  \centering
  \scalebox{.8}{\includegraphics{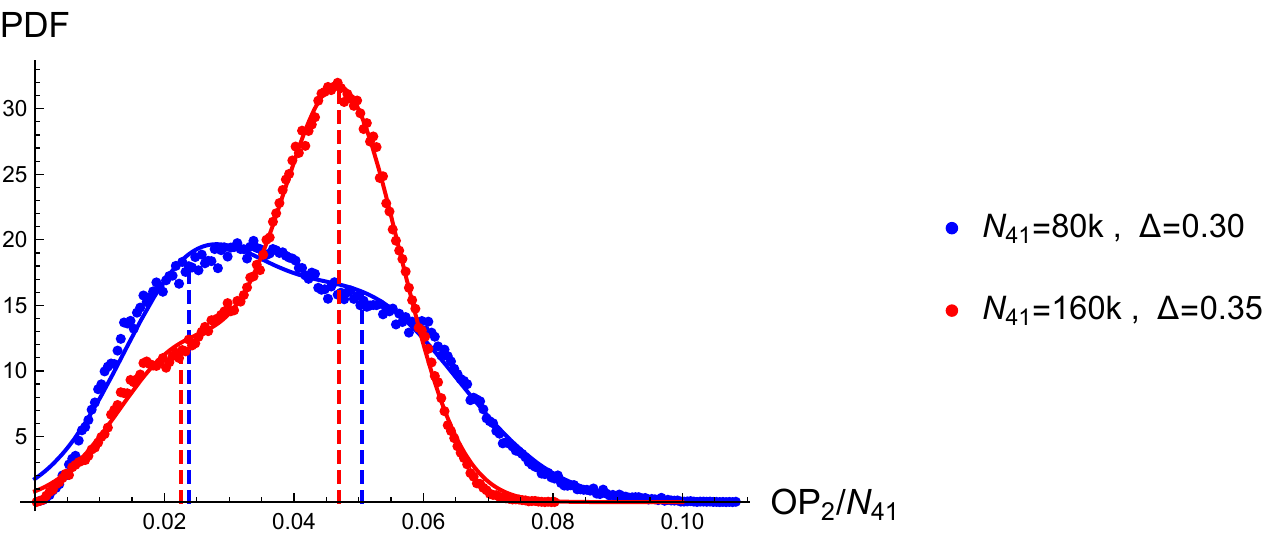}}
  \caption{\small A histogram of the $\text{OP}_2 / N_{4,1}$ order parameter measured at the phase transition point for two different lattice volumes $N_{4,1}=80,000$ (blue) and $N_{4,1}=160,000$ (red). We fit the histogram data to a double Gaussian function (solid line). The position of the two peaks is marked by dashed lines. The peak separation appears to shrink slightly with increasing lattice volume.} 
\label{FigHistogram}
\end{figure}
\begin{figure}[h]
  \centering
  \scalebox{.7}{\includegraphics{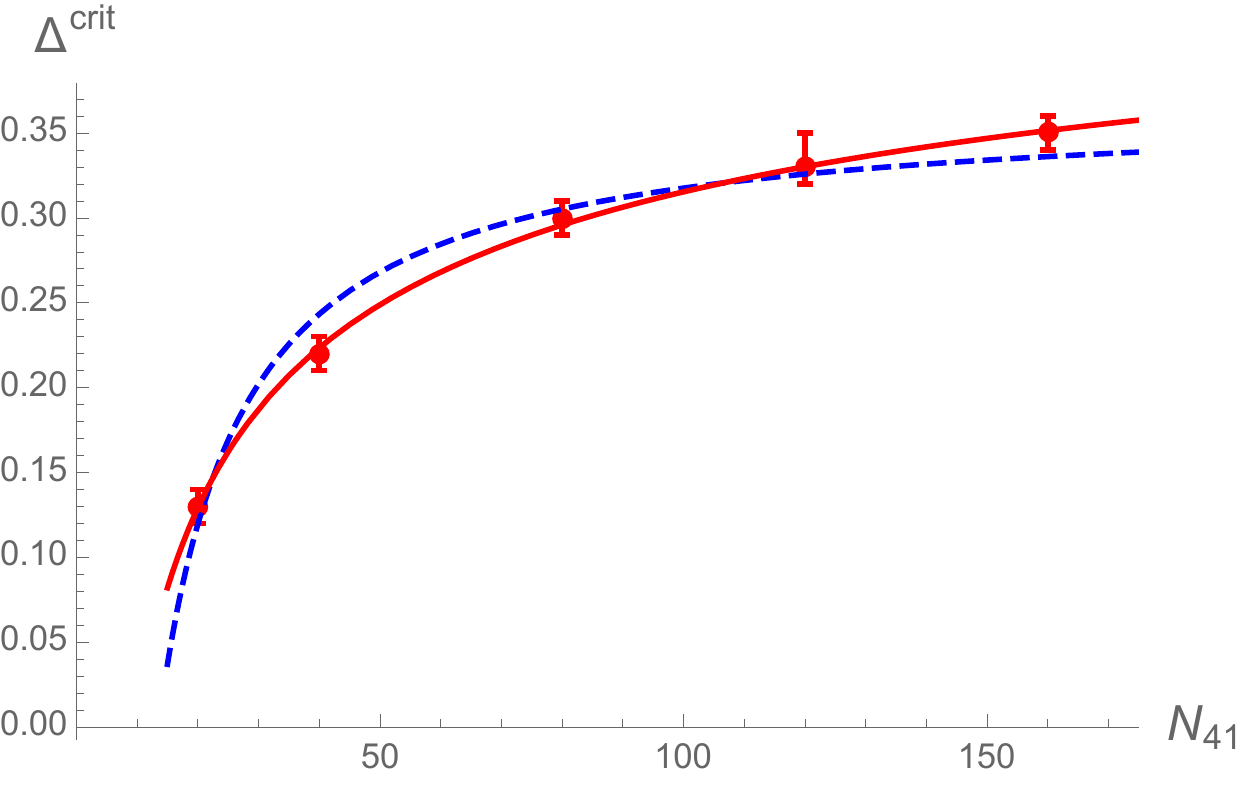}}
  \caption{\small Preliminary results  of C-D phase transition dependence on lattice volume. The 
   (pseudo-)critical points $\Delta^{crit}$  were estimated for fixed $\kappa_0=2.2$ and various total volumes $N_{4,1}$ by looking at peaks in susceptibility $\chi_{\text{OP}_2}$ as described in section \ref{DefOP}. The solid red line corresponds to a fit of Eq.~(\ref{EqVolFit}) to the measured data ($\nu=2.6$), while the dashed blue line uses the same fit but with a critical exponent of $\nu=1$.} 
\label{FigPhaseVol}
\end{figure}

Measuring the behaviour of the order parameter for a number of different lattice volumes will enable us to calculate critical exponents and to analyse the order of the phase transition in detail. This work is still in progress, 
however  preliminary results are promising. In Fig.~\ref{FigPhaseVol}  we plot the position of (pseudo-)critical points $\Delta^{crit}$  as a function of lattice volume $N_{4,1}$.  Using this empirical data we  fit  the function
\begin{equation}\label{EqVolFit}
\Delta^{crit}(N_{4,1})=\Delta^{crit}(\infty)-\alpha \cdot N_{4,1} \- ^{-1/\nu}
\end{equation}
and estimate  the critical exponent $\nu = 2.6 \pm 0.6$ (solid red line in Fig.~\ref{FigPhaseVol}). This value of $\nu$ suggests a continuous transition. For  comparison we also made a fit using a fixed value of $\nu=1$ that would correspond to a first order transition (dashed blue line in Fig.~\ref{FigPhaseVol}), which cannot be completely excluded but appears much less likely. We are currently collecting data at the C-D transition for additional lattice volumes as well as increasing statistics of previous measurements. Unfortunately, this process is computationally very time consuming and  a comprehensive study of the bifurcation transition order will be presented in a separate article.





\end{section}


\section*{Acknowledgements}

The authors wish to acknowledge the support of the grant DEC-2012/06/A/ST2/00389 from the National Science Centre Poland. 


\bibliographystyle{unsrt}
\bibliography{Master}

\begin{thebibliography}{10}

\bibitem{Ambjorn:2008wc}
J.~Ambjorn, A.~Gorlich, J.~Jurkiewicz, and R.~Loll.
\newblock {The Nonperturbative Quantum de Sitter Universe}.
\newblock {\em Phys.Rev.}, D78:063544, 2008.

\bibitem{Ambjorn:2005db}
J.~Ambjorn, J.~Jurkiewicz, and R.~Loll.
\newblock {Spectral dimension of the universe}.
\newblock {\em Phys.Rev.Lett.}, 95:171301, 2005.

\bibitem{Lauscher:2005qz}
O.~Lauscher and M.~Reuter.
\newblock {Fractal spacetime structure in asymptotically safe gravity}.
\newblock {\em JHEP}, 0510:050, 2005.

\bibitem{Horava:2009if}
Petr Ho{\v r}ava.
\newblock {Spectral Dimension of the Universe in Quantum Gravity at a Lifshitz
  Point}.
\newblock {\em Phys.Rev.Lett.}, 102:161301, 2009.

\bibitem{Modesto:2008jz}
Leonardo Modesto.
\newblock {Fractal Structure of Loop Quantum Gravity}.
\newblock {\em Class.Quant.Grav.}, 26:242002, 2009.

\bibitem{Atick:1988si}
Joseph~J. Atick and Edward Witten.
\newblock {The Hagedorn Transition and the Number of Degrees of Freedom of
  String Theory}.
\newblock {\em Nucl.Phys.}, B310:291--334, 1988.

\bibitem{Ambjorn05}
J.~Ambjorn, J.~Jurkiewicz, and R.~Loll.
\newblock {Reconstructing the universe}.
\newblock {\em Phys.Rev.}, D72:064014, 2005.

\bibitem{Catterall:1994pg}
S.~Catterall, John~B. Kogut, and R.~Renken.
\newblock {Phase structure of four-dimensional simplicial quantum gravity}.
\newblock {\em Phys. Lett.}, B328:277--283, 1994.

\bibitem{Bialas:1996wu}
P.~Bialas, Z.~Burda, A.~Krzywicki, and B.~Petersson.
\newblock {Focusing on the fixed point of 4-D simplicial gravity}.
\newblock {\em Nucl.Phys.}, B472:293--308, 1996.

\bibitem{deBakker:1996zx}
Bas~V. de~Bakker.
\newblock {Further evidence that the transition of 4-D dynamical triangulation
  is first order}.
\newblock {\em Phys.Lett.}, B389:238--242, 1996.

\bibitem{Coumbe:2014nea}
Daniel Coumbe and John Laiho.
\newblock {Exploring Euclidean Dynamical Triangulations with a Non-trivial
  Measure Term}.
\newblock {\em JHEP}, 1504:028, 2015.

\bibitem{Ambjorn:2013eha}
J.~Ambjorn, L.~Glaser, A.~Goerlich, and J.~Jurkiewicz.
\newblock {Euclidian 4d quantum gravity with a non-trivial measure term}.
\newblock {\em JHEP}, 1310:100, 2013.

\bibitem{Regge:1961px}
T.~Regge.
\newblock {General Relativity Without Coordinates}.
\newblock {\em Nuovo Cim.}, 19:558--571, 1961.

\bibitem{Ambjorn:2011cg}
J.~Ambjorn, S.~Jordan, J.~Jurkiewicz, and R.~Loll.
\newblock {A Second-order phase transition in CDT}.
\newblock {\em Phys.Rev.Lett.}, 107:211303, 2011.

\bibitem{Ambjorn:2012ij}
Jan Ambjorn, S.~Jordan, J.~Jurkiewicz, and R.~Loll.
\newblock {Second- and First-Order Phase Transitions in CDT}.
\newblock {\em Phys.Rev.}, D85:124044, 2012.

\bibitem{Ambjorn:2015qja}
Jan Ambjørn, Daniel~N. Coumbe, Jakub Gizbert-Studnicki, and Jerzy Jurkiewicz.
\newblock {Signature Change of the Metric in CDT Quantum Gravity?}
\newblock {\em JHEP}, 08:033, 2015.

\bibitem{Ambjorn:2014mra}
Jan Ambjørn, Jakub Gizbert-Studnicki, Andrzej Görlich, and Jerzy Jurkiewicz.
\newblock {The effective action in 4-dim CDT. The transfer matrix approach}.
\newblock {\em JHEP}, 1406:034, 2014.

\end{thebibliography}


\end{document}